\documentclass[aps,prl, twocolumn,superscriptaddress,groupedaddress]{revtex4}
\usepackage{amssymb}
\usepackage{amsmath}
\usepackage{graphicx}
\usepackage{epstopdf}
\usepackage{bm}% bold math
\usepackage{gensymb}
\usepackage{soul}
\usepackage{sidecap}
\usepackage[normalem]{ulem}
\usepackage{ mathrsfs }
\usepackage {times}
\usepackage{float}
\usepackage[caption = false]{subfig}
\usepackage{subfig}
\usepackage{enumerate}
\usepackage{multirow}
\usepackage{tabularx}
\usepackage{array}
\usepackage{slantsc}
\usepackage{lmodern}
\newcommand\redout{\bgroup\markoverwith{\textcolor{red}{\rule[.5ex]{2pt}{0.4pt}}}\ULon}

\usepackage[colorlinks=true,citecolor=blue]{hyperref}
\hypersetup{colorlinks=true,citecolor=blue,linkcolor=red,urlcolor=blue}

\usepackage{hyperref,cleveref}
\graphicspath{{.}{./Figs/}}
\usepackage{color}
\usepackage[dvipsnames]{xcolor}
\usepackage{xcolor}
\begin{document}
\title{Integrable Quantum Many-Body Sensors for AC Field Sensing}

%\title{Retrieving Heisenberg limit in partially accessible many-body systems}

\author{Utkarsh Mishra}
\author{Abolfazl Bayat}
\email{abolfazl.bayat@uestc.edu.cn}
\affiliation{Institute of Fundamental and Frontier Sciences, University of Electronic Science and Technology of China, Chengdu 610051, China}
\begin{abstract}
Quantum sensing is inevitably an elegant example of the supremacy of quantum technologies over their classical counterparts.  One of the desired  endeavors of quantum metrology is AC field sensing. Here, by means of analytical and numerical analysis, we show that integrable many-body systems  can be exploited efficiently for detecting the amplitude of an AC field. Unlike the conventional strategies in using the ground states in critical many-body probes for parameter estimation, we only consider partial access to a subsystem.  Due  to the periodicity of the dynamics,  any local block of the system saturates to a steady state which allows achieving sensing precision well beyond the classical limit, almost reaching the Heisenberg bound. We associate the enhanced quantum precision to closing of the Floquet gap,  resembling the features of quantum sensing in the ground state of critical systems. 
We show that the proposed protocol can also be realized in near-term quantum simulators, e.g. ion-traps, with a limited number of qubits. We show that in such systems a simple block magnetization measurement and a Bayesian inference estimator can achieve very high precision AC field sensing.
\end{abstract}

\flushbottom
\maketitle
\thispagestyle{empty}

%\noindent Please note: Abbreviations should be introduced at the first mention in the main text – no abbreviations lists. Suggested structure of main text (not enforced) is provided below.

\section*{Introduction}

Quantum systems have emerged as excellent sensors for detecting various types of fields~\cite{RevModPhys.89.035002}, including weak magnetic~\cite{Kominis:2003ab,Budker:2007aa,PhysRevLett.98.200801,Taylor:2008aa,IEEESensors,Bal:2012aa}, electric~\cite{PhysRevLett.95.230801,Morello:2010aa,Sedlacek:2012aa,RevModPhys.87.1419,Fan_2015}, and gravitational fields~\cite{Schnabel2010}, due to their extreme sensitivity against variation in the environment.  The prospect of applications for quantum sensing is very wide covering material science~\cite{Lovchinsky503} to biomedical analysis~\cite{Jensen:2016aa,McGuinness:2011aa}.  In particular, AC field sensing has been the subject of intense theoretical and experimental research  for the estimation of amplitude~\cite{Timoney2011,Baumgart2016,Weidt2016}, frequency~\cite{Khodjasteh2005,Lang_2015}, and phase~\cite{Shora,PhysRevLett.106.080802,Kominis:2003aa,PhysRevLett.103.220802,PhysRevLett.106.030802, Rondin_2014,doi:10.1146/annurev-physchem-040513-103659,PARIS2009}. The majority of these protocols, mainly implemented in nitrogen vacancy centers, utilize a series of spin-echo pulses to accumulate the information about the AC field in the phase of a coherent superposition of a single qubit, which is then converted into the amplitude at the readout stage~\cite{Abraham1990,Mueller2014, Hansom2014}. However, the ultimate precision is limited by the number of spin-echo pulses that one can apply within the coherence time.  To enhance the precision, one can increase the number of particles, although,  once the particles start to interact, the precision is severely hindered~\cite{PhysRevB.80.115202}.  In Ref.~\cite{2019arXiv190710066Z}, a complex pulse structure has been designed to suppress the interaction between the particles and enhance the sensing precision.  
Therefore, an important open question is whether one can go beyond the spin-echo procedure and harness the interaction between particles, instead of suppressing it, for AC field sensing.

%%%%%%%%%%%%%%%%%%%%% FIG1 %%%%%%%%%%%%%%%%%%%%%%%%%%%%%%%%%%%%%%%%%%%%%%
\begin{figure}[h!]
\centering
\includegraphics[height=0.24\textheight]{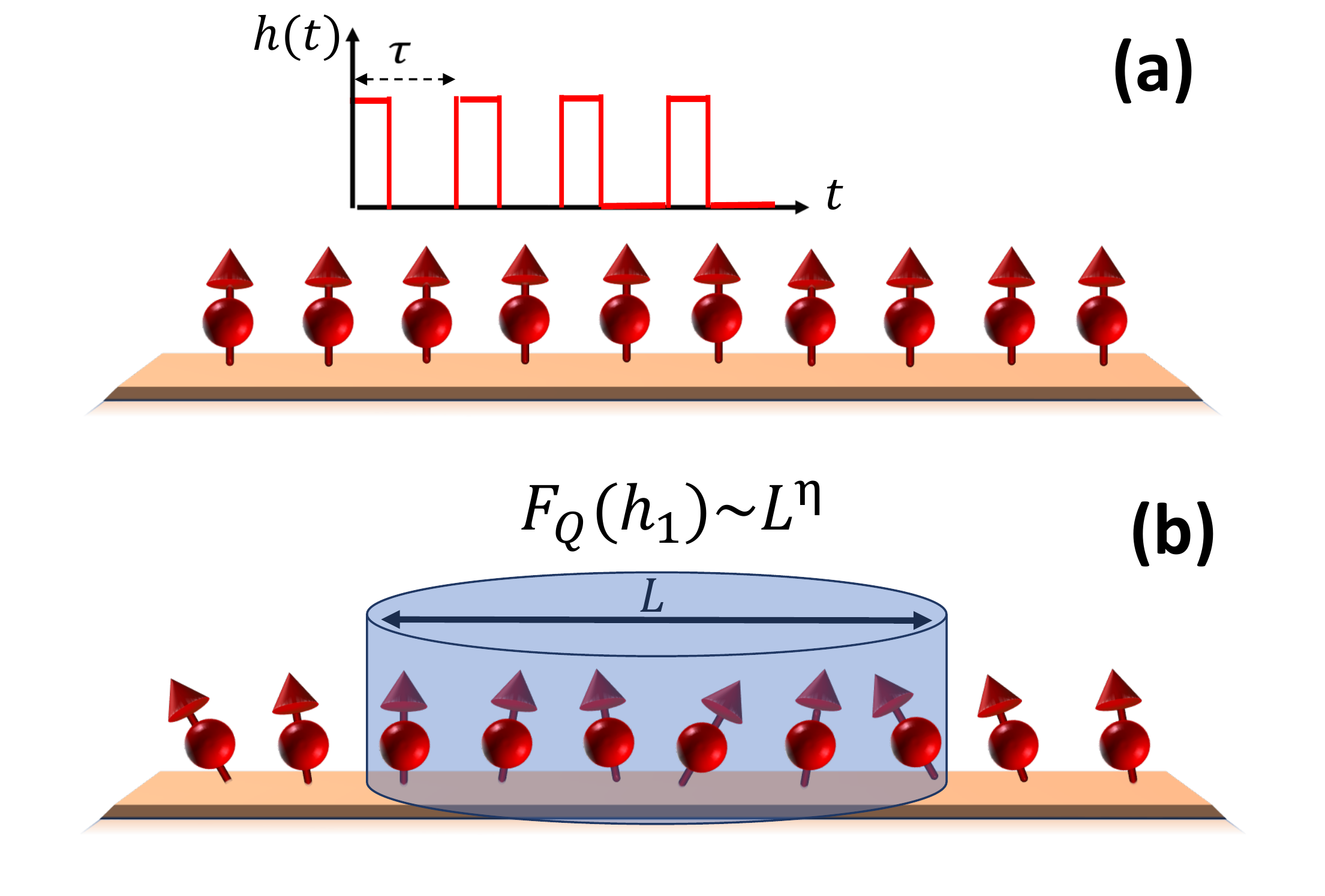}
\caption{ (a) The many-body quantum system  of spin-1/2 particles,  prepared in its ground state, is interacting with a time-periodic magnetic field, $h(t)$, of period $\tau$ and strength $h_1$. (b) In the steady state, a block of $L$ contiguous spins are measured resulting in a quantum Fisher information  which scales with $L$ as $F_Q(h_1)\sim L^{\eta}$.  
}
\label{fig:fig1}
\end{figure}

The quality of any sensing protocol, either classical or quantum, is quantified by the uncertainty in the estimation of an unknown parameter $h$ which is fundamentally bounded by the Cram\'er-Rao inequality as $\mbox{Var}(h)\geq 1/F$~\cite{Braunstein1994}. Here, $\mbox{Var}(h)$ is the variance of the estimation with respect to an unbiased estimator, $F\sim L^\eta$ is the  Fisher information,  $L$ is the number of resources, and $\eta$ is a positive constant~(See Refs.~\cite{Liu_2019,Review_QFI} for a recent review on quantum Fisher information). Classical systems, at best, can result in  $\eta = 1$ known as the standard limit. By harnessing quantum entanglement, e.g. in the specific form of GHZ 
~\cite{Giovannetti2006} and N00N~\cite{Cable2007} states, one can enhance the sensitivity to $\eta=2$, known as the Heisenberg limit. However, these states are extremely sensitive to decoherence and particle loss~\cite{Duer2000,kolodynski2013efficient} making them impractical for real applications. In addition, any interaction between the particles deteriorates the sensing quality~\cite{PhysRevA88052117}.  
One can also  exceed the standard limit through adaptive~\cite{bonato2016optimized,said2011nanoscale,higgins2007entanglement, berry2009perform,higgins2009demonstrating,danilin2018quantum,shlyakhov2018quantum} or continuous measurements~\cite{gammelmark2014fisher} using single particle sensors. 

While in the GHZ-based quantum sensing, the interaction between particles should be avoided, in a fundamentally different route, one can harness the interaction in strongly correlated many-body quantum systems in~\cite{zanardi2008quantum,invernizzi2008optimal,
salvatori2014quantum,bina2016dicke,boyajian2016compressed,
rams2018limits,mehboudi2016achieving} and out~\cite{SegioPRA2008,SegioPRL2008,SegioPRA2009,SegioPRA2010,kiukas2017remote,jones2020remote} of equilibrium for sensing. In fact, thanks to the emergent of multipartite entanglement~\cite{guhne2005multipartite,guhne2006energy,campbell2010multipartite,giampaolo2013genuine,giampaolo2014genuine,bayat2017scaling}, many-body systems near criticality provide enhanced quantum precision of $\eta = 2/\nu$~\cite{zanardi2008quantum,invernizzi2008optimal,salvatori2014quantum,bina2016dicke,boyajian2016compressed,rams2018limits}, where $\nu$ is the critical exponent in charge of the divergence of correlation length~\cite{Duttaa,Sachdev2009}. %Interestingly, for $\nu<1$, many-body system can even provide super-Heisenberg scaling~\cite{rams2018limits,Gong2008,Greschner2013}.
In addition, the evolution of many-body systems has also been used for sensing local~\cite{jones2020remote} and global~\cite{Raghunandan2018} DC fields as well as extracting information about the spectral structure of time-varying fields~\cite{roushan2017spectroscopic,bayat2018measurement,bayat2015universal}. In most of these works, either static or dynamic, it is dominantly assumed that the whole system is accessible for measurement which may not be practical. Nonetheless, quantum enhancement in many-body sensors with only partial access to a subsystem has hardly been explored and it is not clear whether criticality can still enhance the precision in such scenarios. One may raise question whether strongly correlated many-body systems can also be beneficial for AC field sensing. If so, do they provide precision beyond the standard limit? What would happen if only partial access to a subsystem is available? The importance lies in the fact that  the AC field excites high energy eigenstates and thus  the notion of ground state criticality will no longer exist and thus a new theory is needed. Recently, we have shown that one can gain quantum enhanced sensing in periodically driven systems  even with the partial accessibility for sensing DC magnetic fields~\cite{oursensing}. Here, we generalize this approach for sensing AC magnetic fields. In the present work, we also demonstrate the practicality of the  sensing protocol using Bayesian inference. 
%Broadly speaking, for the Hamiltonians consists of on-site local terms, the resultant uncertainty in the magnetic field achieve Heisenberg limit ($1/N$)~\cite{}.

The main findings of the paper are: (i) the quantum Fisher information of a block, with respect to amplitude of the AC magnetic field, peaks along a line, making it far more versatile than the critical systems; (ii) the line of the peak of the quantum Fisher information coincides with the line of vanishing Floquet gap; and (iii) at the closing of the Floquet gap, the quantum Fisher information scales well beyond the standard limit, shows quantum enhanced sensing. Finally, we numerically analyze the proposal for an ion trap.
%we show that our protocol can be used in ion traps.

\section*{Results}

%\section{Results}
\section {The setup for sensing} 
We consider an interacting spin-1/2 Ising chain of length $N$ in a transverse field to serve as a many-body probe for sensing a time-periodic magnetic field, $h(t)$, which is assumed to be along the transverse direction.  The Hamiltonian of the model is written as
\begin{eqnarray}
H(t)=- J \sum_{i=1}^{N}\hat{\sigma}^{x}_{i}\hat{\sigma}^{x}_{i+1} - \sum_{i}(h_0 + h(t))\hat{\sigma}^{z}_{i},
\label{eq:model}
\end{eqnarray} 
where, $J>0$ is the nearest-neighbor spin-spin interaction, $h_0$ is a DC external magnetic field which is tunable,  $\hat{\sigma}_{i}^{x/z}$  are Pauli matrices at site $i$, and the periodic boundary conditions is assumed, i.e., $\hat{\sigma}_{N+1}^{x/z}=\hat{\sigma}_{1}^{x/z}$.  
In the absence of $h(t)$, the Hamiltonian in Eq.~(\ref{eq:model}) is known to  exhibit a quantum phase transition at $h_0=h_c$ such that $h_c/J=1$.     
The time-dependent field $h(t)$, as we will show later, can be any periodic function with a nonzero mean over a period, such as  Dirac delta-kick or square pulses. A schematic picture of the system is given in Fig.~\ref{fig:fig1}. To begin with, the time-dependent form of the magnetic field, $h(t)$, is taken in the form of a Dirac delta-kick  as
\begin{eqnarray}
h(t) = h_1\sum\limits_{n=0}^{n = \infty}\delta(t-n\tau),
\label{eq:dirac}
\end{eqnarray}
where, the strength of the kick is $h_1$ whose estimation will be investigated in this paper. The above Hamiltonian in the presence of $h(t)$, Eq.~(\ref{eq:dirac}), is time periodic, i.e.,  $H(t){=}H(t+n\tau)$ with $\tau$ being the time period, which is known a priory, and $n$ being integer valued. The time evolution  monitored in steps of $t{=}n\tau$ is referred as \textit{stroboscopic} in the literature~\cite{DAlessio2014,Lazarides2014}. The initial state of the evolution is taken to be a fully polarized state where each spins are in the eigenbasis of $\hat{\sigma}_z$ with eigenvalue $+1$, i.e.,      $|\Psi(0)\rangle${=}$|\uparrow \otimes \uparrow \otimes\ldots\otimes\uparrow\rangle.$ The role of other initial states is discussed in more detail later. 
The time evolved state of the system is  $|\Psi(t)\rangle{=}U(t,0)|\Psi(0)\rangle$, where 
\begin{equation}\label{eq:Evolution_U}
U(t,0){=}{\cal T}e^{-i \int_{0}^{t} H(t)dt},
\end{equation}
with ${\cal T}$ being the time order operator. For such a case, the subsequent dynamics can be obtained from the knowledge of one time period propagator $U(\tau,0)$ and is termed as Floquet operator. The Floquet evolution has already been found useful in explaining the emergence of thermal states under periodic driving~\cite{Lazarides2014}, engineering exotic topological phases of matter~\cite{Thakurathi2013}, dynamically decoupling the interaction between the particles~\cite{Lang_2015, 2019arXiv190710066Z} and efficiently being simulated on digital quantum simulators~\cite{Sieberer2019}.

The Hamiltonian in Eq.~(\ref{eq:model}) can be solved exactly using Jordan-Wigner transformation (JW), as elaborated in~\cite{Lieb1961,Barouch1971}. 
We outline the key steps and present detailed calculations in the  Sec. A of the Supplementary Materials (SM). The first step is to map the spin operators, $\hat{\sigma}_{i}$, into fermionic operators, $\hat{c}^{\dagger}_{i}(\hat{c_{i}})$, via the JW transformations:
\begin{eqnarray}
\hat{\sigma}^{-}_{j} &=& e^{i\pi\sum_{i=1}^{j-1}\hat{\sigma}^{+}_{i}\hat{\sigma}^{-}_{i}}c_{j}\nonumber\\
\hat{\sigma}^{+}_{j} &=& c^{\dagger}_{j}e^{-i\pi\sum_{i=1}^{j-1}\hat{\sigma}^{+}_{i}\hat{\sigma}^{-}_{i}},
\label{eq:JW}
\end{eqnarray}
where, $\hat{\sigma}^{\pm}_{j} = (\hat{\sigma}^{x}_j\pm\hat{\sigma}^{y}_j)/2 $. By defining Fourier space fermionic operator as $d_k = \frac{1}{\sqrt{N}}\sum_{j}e^{i k j} c_j$, one gets $H(t)=\sum_{k}H_{k}$, where  $H_{k}$ being the Hamiltonian of the $k^{th}$ subspace given by $H_k = (h(t)+J\cos(k))(d^{\dagger}_{k}d_{k}-d_{-k}d^{\dagger}_{-k}) + J\sin(k)(d^{\dagger}d^{\dagger}_{-k}-d_{k}d_{-k})$. The time-evolved state can be obtained using Eq.~(\ref{eq:Evolution_U}) and the fact that the Hamiltonian is a sum of independent modes, $k$, as
\begin{eqnarray}
|\Psi(t=n\tau)\rangle &= &[U(\tau,0)]^n|\Psi(0)\rangle=e^{-i n H^F \tau}|\Psi(0)\rangle
\nonumber\\
 &=&\otimes_{k>0} e^{-inH^{F}_{k}\tau}|\psi^{0}_{k}\rangle.
\end{eqnarray}
Here, $|\Psi(0)\rangle = \otimes_{k}|\psi^{0}_{k}\rangle$ is the initial state and $H^{F}_{k}$ is termed as Floquet Hamiltonian. The Floquet Hamiltonian $H^{F}_{k}$ turns out to be simple to obtain for the delta-kick field. If, over a period $\tau$, the initial and the final Hamiltonians are $H^{i}_{k}$ and $H^{f}_{k}$, respectively, then 
\begin{eqnarray}
H^{F}_{k} = |\vec{{\cal \mu}}^{F}_{k}|\hat{n}^{F}_{k}.\vec{\sigma}_p,
\label{eq:HFk}
\end{eqnarray} 
where, $\vec{\sigma}_p=(\hat{\sigma}^{x}_p,\hat{\sigma}^{y}_p,\hat{\sigma}^z_p)$ are the pseudospin-1/2 operators, $\hat{n}^{F}_{k} = \vec{{\cal \mu}}^F_{k}/|\vec{{\cal \mu}}^{F}_{k}|$, and the Floquet quasi energies $|\vec{{\cal \mu}}^{F}_{k}|$ are given by
\begin{eqnarray}
|\vec{{\cal \mu}}^{F}_{k}| &= &\frac{1}{\tau}\cos^{-1}\Big[\cos (|{\vec{\cal \mu}}^{i}_{k}| \tau) \cos(|\vec{{\cal \mu}}^{f}_{k}|\tau) \nonumber\\
&-& \hat{n}^{i}_{k}.\hat{n}^{f}_{k}\sin (|\vec{{\cal \mu}}^{i}_{k}|\tau)\sin (|\vec{{\cal \mu}}^{f}_{k}|\tau)\Big],
\label{eq:kfloquet}
\end{eqnarray}
and
\begin{eqnarray}
\hat{n}^{F}_{k}  &= &\frac{1}{\sin(|\vec{{\cal \mu}}^{F}_{k}|\tau)}\Big[\hat{n}^{i}_{k} \sin(|\vec{{\cal \mu}}^{i}_{k}|\tau)\cos(|\vec{{\cal \mu}}^{f}_{k}|\tau)\nonumber\\
&+&\hat{n}^{f}_{k} \sin(|\vec{{\cal \mu}}^{f}_{k}|\tau)\cos(|\vec{{\cal \mu}}^{i}_{k}|\tau) \nonumber\\
&-&\hat{n}^{i}_{k}\times \hat{n}^{f}_{k} \sin(|\vec{{\cal \mu}}^{i}_{k}|\tau)\sin(|\vec{{\cal \mu}}^{f}_{k}|\tau)\Big],
\end{eqnarray}
where, $H^{i}_{k} = |\vec{{\cal \mu}}^{i}_{k}|\hat{n}^{i}_{k}.\vec{\sigma}_p$  and similarly for $H^{f}_{k}$ with $\hat{n}^{i(f)}_{k} = \vec{{\cal \mu}}^{i(f)}_{k}/|\vec{{\cal \mu}}^{i(f)}_{k}|$. For the  delta-kick magnetic field,  $\vec{{\cal \mu}}^{i}_{k} = (0, J\sin(k), h_0+J\cos(k))$, $\vec{{\cal \mu}}^{f}_{k} = (0, 0, h_1)$, and the Floquet Hamiltonian, obtained in Eq.~(\ref{eq:HFk}), is a $2\times 2$ matrix.  

For a given wave function $|\Psi(t)\rangle$ of a many-body quantum system, partial accessibility on a length scale $L\ll N$ is well described by a reduced density matrix, $\rho_L$, which is given by
\begin{equation}
\rho_L(t) = \mbox{Tr}_{N-L}|\Psi(t)\rangle \langle \Psi(t)|,
\label{eq:reduced_dm}
\end{equation} 
where  $\mbox{Tr}_{N-L}$ stands for the partial trace overall sites except the spins within the block $L$. It is worth emphasizing that although the density matrix of the full system, given by $\rho(t){=}|\Psi(t)\rangle\langle \Psi(t)|$, is pure, the density matrix $\rho_L(t)$ is mixed as the state $|\Psi(t)\rangle$ gets more entangled with increasing $t$. Thanks to the periodic boundary condition, the choice of the location of the block is irrelevant and only its size $L$ is important.  As the system evolves, the information of $h_1$ is imprinted on the quantum state $\rho_L(t)$ which can be extracted by
performing proper measurements and feeding the results
into an estimator algorithm, such as Bayesian inference (the details are presented in Sec. VII). In the long-time, as we will see in the following sections, the dynamics of the observables associated with $\rho_L$ equilibrate to a steady state value, which will be incorporated into our sensing protocol for estimating $h_1$. One can properly tune the DC field $h_0$, as an extra controllable parameter, to enhance the sensitivity of the system to the variation of $h_1$.  Moreover,  without loss of generality, we fix the time-period $\tau$ to be $J\tau{=}0.2$ as,  we will see that, for all $J\tau{\leq}1$ the local steady state can be used for parameter estimation.

 \section {Estimation theory} 
We, in this section,  review the quantum estimation theory for inferring an unknown parameter encoded in a general density matrix.  Any estimation protocol relies on two crucial ingredients: (i) a measurement setup that measures the system on a specific basis and (ii) an estimator algorithm that uses the measured data for inferring the value of the unknown parameter. 
The precision of estimating the unknown parameter, $h_1$,  quantified by the statistical variance, is bounded by the Cram\'er-Rao inequalities~\cite{Braunstein1994,PARIS2009}
\begin{equation}
\mbox{Var}(h_1)\geq \frac{1}{M F_C(h_1)}\geq \frac{1}{M F_Q(h_1)},
\label{eq:CRB}
\end{equation} 
where, $M$ is the number of samples, $F_C$ and $F_Q$ are the classical and quantum Fisher information,  respectively. The above inequalities show that the variance of any \textit{unbiased} estimator of a parameter cannot be lower than the inverse of the Fisher information.
When the measurement basis is fixed, say by a set of positive valued measurements (POVM) $\{\Pi_r\}$, the above inequality is bounded by the classical Fisher Information (CFI) $F_C$, which is also known as the classical Cram\'er-Rao inequality.   In this case, the equality is achieved when the estimator algorithm is optimized.  The classical Fisher information is given by
\begin{eqnarray}
F_C(h_1)=\sum_{r} \frac{(\partial_{h_1} p_{r})^2}{p_r},
\label{eq:CFI}
\end{eqnarray}
where, $p_{r}(h_1){=}\mbox{Tr}[\rho_L (h_1) \Pi_{r}]$ is the probability of obtaining the outcome $r$ and $\partial_{h_1}p_{r}{=}\frac{\partial p_r}{\partial h_1}$.  Since the POVM satisfies $\sum_{r}\Pi^{\dagger}_{r}\Pi_{r}{=}{\mathbb I}$, where ${\mathbb I}$ is the identity matrix in the state space, it automatically implies that $\sum_{r} p_{r}{=} 1$. One can further tighten the classical Cram\'er-Rao inequality by optimizing the measurement basis over all possible POVMs which then results in a new bound, given by the quantum Fisher information (QFI) $F_Q$, as stated in Eq.~(\ref{eq:CRB}). In this case, the inequality is called the quantum Cram\'er-Rao inequality.   Note that the QFI is independent of the measurement basis and the equality is achieved when both estimation algorithm and measurement basis are chosen to be optimal. 

\noindent
For the density matrix,  $\rho_L$,  the QFI  is given by~\cite{PARIS2009} 
\begin{equation}
F_{Q}=\sum_{r,s=1}^{2L}\frac{2 {\Re}(\langle \lambda_{r}|\partial_{h_1}\rho_{L}|\lambda_{s}\rangle \langle \lambda_{s}|\partial_{h_1}\rho_{L}|\lambda_{r}\rangle)}{\lambda_{r}+\lambda_s},
\label{eq:QFI} 
\end{equation}
where, $\rho_L{=}\sum_{r=1}^{2L}\lambda_{r}|\lambda_{r}\rangle \langle \lambda_{r}|$ is the spectral decomposition of $\rho_L$ with $\lambda_r$ and $|\lambda_{r}\rangle$ being the eigenvalues and eigenvectors, respectively. ${\Re}[\cdot]$ denotes the real part and the sum in Eq.~(\ref{eq:QFI}) excludes terms for which $\lambda_r+\lambda_s{=}0$. 
%Note that, since the measurement has been optimized, the QFI is a measurement independent quantity which implies that $F_Q\geq F_C$. 
The computation of the time-dependent QFI of $\rho_L$ at time $t$ is explained in Sec. B of SM.

\begin{figure}
\centering
\includegraphics[height=0.25\textheight]{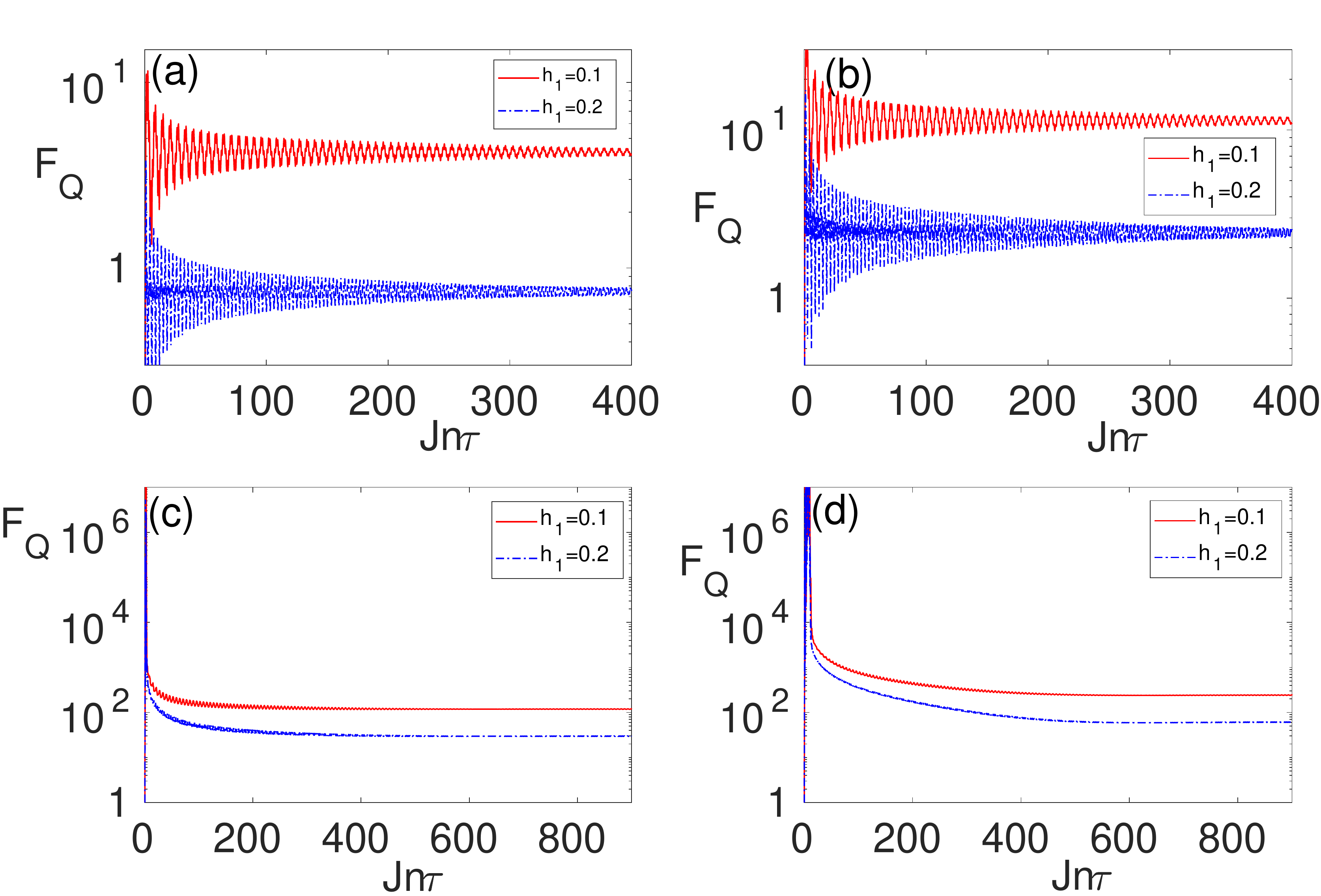}
\caption{The time-evolution of quantum Fisher information $F_{Q}$ as a function of time $t{=}n\tau$ for different values of $h_1/J{=}0.1$ (regular red line), $h_1/J{=}0.2$ (dashed dotted blue line) and various block sizes: (a) $L{=}1$; (b) $L{=}2$; (c) $L{=}10$; and (d) $L{=}20$. 
}
\label{fig:fig2}
\end{figure}

%%%%%%%%%%%%%%%%%%%%% FIG3%%%%%%%%%%%%%%%%%%%%%%%%%%%%%%%%%%%%%%%%%%%%%%

\begin{figure}
\centering
\includegraphics[height=0.3\textheight]{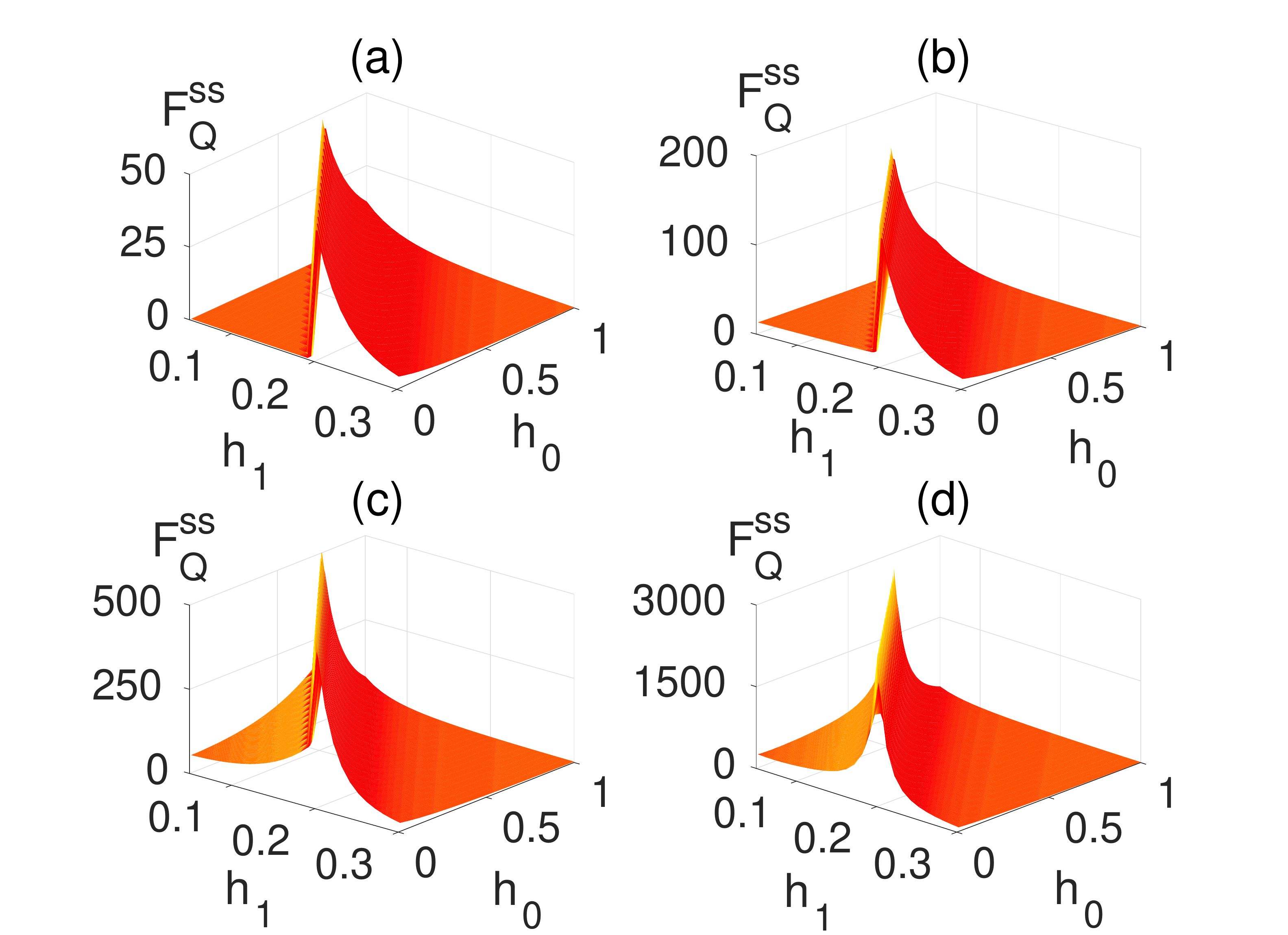}
\caption{Variation of average long-time quantum Fisher information $F_Q^{ss}$ with respect to $h_1$ and $h_0$ for different block sizes: (a). $L{=}1$; (b) $L{=}2$; (c) $L{=}4$; and (d) $L{=}10$.  For the numerical calculation of quantum Fisher information, we choose $dh_1{=}10^{-3}$, see Eq.~(\ref{eq:QFI}).  Here $N=2000$ and $J\tau=0.2$.
}
\label{fig:fig3}
\end{figure}

\section {Quantum Fisher information analysis}
To quantify the sensitivity of our probe for inferring $h_1$, one can use the QFI of $\rho_L$ for different block sizes.  
In Figs.~\ref{fig:fig2}(a)-(d), we plot the dynamics of QFI, $F_{Q}(t)$,  as a function of time $t{=}n\tau$, for different values of $h_1$ when $h_0$ is tuned at $h_0/J{=}1$. Each panel in Figs.~\ref{fig:fig2}(a)-(d) represents a different block size namely: (a) $L{=}1$, (b) $L{=}2$, (c) $L{=}4$,  and (d)  $L{=}10$.  
The QFI shows oscillatory behavior with damping amplitudes which at long times saturates to a steady state value depending on $h_1$.  The steady state QFI value becomes significantly larger as the block size $L$ increases, implying that the sensing  precision considerably enhances as $L$ increases. The long-time oscillations in the QFI persist because of the finite total system size $N$. To obtain the steady state value, we consider time averaged QFI given by
\begin{eqnarray}
F^{ss}_Q = \frac{1}{\tau(n_{max}-n_{min})}\sum_{t = n_{min}\tau}^{n_{max}\tau}F_{Q}(t).
\label{eq:ssfisher}
\end{eqnarray}
Typically, for our numerical calculation, $n_{min}$ and $n_{max}$ are taken to be $ 4000$ and $4400$, respectively for Figs.~\ref{fig:fig2}. These values are chosen to include a few oscillations of $F_{Q}(t)$. Once this condition is satisfied, any further widening of the range of $n_{min}$ and $n_{max}$ will give almost the same value of $F^{ss}_{Q}$. In fact, one can take the limit $n\to \infty$ and obtained $F^{ss}_{Q}{=}\lim\limits_{n\rightarrow \infty} F_Q(n\tau)$ for the state $\rho_L$. To do this, we note that
\begin{eqnarray}
U_{k}(n\tau) {=}e^{-i\mu_{k}^{F,+}n\tau}|\mu^{F,+}_{k}\rangle\langle\mu^{F,+}_{k}|+e^{-i\mu_{k}^{F,-}n\tau}|\mu^{F,-}_{k}\rangle\langle\mu^{F,-}_{k}|,
\end{eqnarray}
where, $\mu_{k}^{F,\pm},|\mu^{F,\pm}_{k}\rangle$ are the eigenvalues and eigenvectors of the Floquet Hamiltonian, respectively. Then the expectation value of  ${\cal C}_{i,j}(n\tau){=}\langle \Psi (n\tau)| c^{\dagger}_{i}c_{j}|\Psi(n\tau)\rangle$ and ${\cal I}_{i,j}(n\tau){=} \langle \Psi(n\tau)| c^{\dagger}_{i}c^{\dagger}_{j}|\Psi(t)\rangle$ between the fermionic operators ($i,j=1,\ldots,L$) can be obtained as
\begin{eqnarray}
{\cal C}_{i,j}(n\tau)&=&\frac{2}{N}\sum_{k>0}\cos(k(i-j))\langle \psi^{0}_{k}| U^{\dagger}_{k}(n\tau)d^{\dagger}_{k}d_{k}U_{k}(n\tau) |\psi^{0}_{k} \rangle\nonumber\\
&=& \frac{2}{N}\sum_{k>0}\cos(k(i-j))\Big[r^{+}_kr^{*+}_k\langle \mu^{F,+}_k|d^{\dagger}_{k}d_{k}|\mu^{F,+}_k\rangle\nonumber\\
&+&r^{-}_kr^{*-}_k\langle \mu^{F,-}_k|d^{\dagger}_{k}d_{k}|\mu^{F,-}_k\rangle\nonumber\\
&+& e^{i(\mu^{F,+}_k-\mu^{F,-}_k)n\tau}r^{+}_kr^{*-}_k\langle\mu^{F,+}_k|d^{\dagger}_{k}d_{k}|\mu^{F,-}_k\rangle\nonumber\\
&+& e^{i(\mu^{F,-}_k-\mu^{F,+}_k)n\tau}r^{-}_kr^{*+}_k\langle\mu^{F,-}_k|d^{\dagger}_{k}d_{k}|\mu^{F,+}_k\rangle
\Big].
\label{eq:Cij}
\end{eqnarray}

\begin{eqnarray}
{\cal I}_{i,j}(t)&=&\frac{2i}{N}\sum_{k>0}\sin(k(i-j))\langle \psi^{0}| U^{\dagger}_{k}(n\tau)d^{\dagger}_{k}d^{\dagger}_{-k}U_{k}()n\tau |\psi^{0} \rangle\nonumber\\
&=& \frac{2i}{N}\sum_{k>0}\sin(k(i-j))\Big[r^{+}_kr^{*+}_k\langle \mu^{F,+}_k|d^{\dagger}_{k}d^{\dagger}_{-k}|\mu^{F,+}_k\rangle\nonumber\\
&+&r^{-}_kr^{*-}_k\langle \mu^{F,-}_k|d^{\dagger}_{k}d^{\dagger}_{-k}|\mu^{F,-}_k\rangle\nonumber\\
&+& e^{i(\mu^{F,+}_k-\mu^{F,-}_k)n\tau}r^{+}_kr^{*-}_k\langle \mu^{F,+}_k|d^{\dagger}_{k}d^{\dagger}_{-k}|\mu^{F,-}_k\rangle\nonumber\\
&+& e^{i(\mu^{F,-}_k-\mu^{F,+}_k)n\tau}r^{-}_kr^{*+}_k\langle\mu^{F,-}_k|d^{\dagger}_{k}d^{\dagger}_{-k}|\mu^{F,+}_k\rangle
\Big].
\label{eq:Iij}
\end{eqnarray}
Here,  $r_{k}^{\pm}=\langle \psi^{0}_{k}|\mu^{F,\pm}_{k}\rangle$, describes the overlap of the initial state with that of the Floquet eigenstates.  Taking the limit  $n\to \infty$ and $N\to \infty$, it is compatible to drop the fast oscillating cross-term from Eqs.~(\ref{eq:Cij}-\ref{eq:Iij}). Thus, we obtain the correlation functions in the steady-state as
\begin{eqnarray}
{\cal C}^{\infty}_{i,j} &=&\frac{1}{\pi}\int_{0}^{\pi}dk\cos(k(i-j)) \Big[|r^{+}_k|^2\langle \mu^{F,+}_k|d^{\dagger}_{k}d_{k}|\mu^{F,+}_k\rangle\nonumber\\
&+&|r^{-}_k|^{2}\langle \mu^{F,-}_k|d^{\dagger}_{k}d_{k}|\mu^{F,-}_k\rangle\Big],
\end{eqnarray}
\begin{eqnarray}
{\cal I}^{\infty}_{i,j} &=&\frac{i}{\pi}\int_{0}^{\pi}dk\sin(k(i-j)) \Big[|r^{+}_k|^2\langle \mu^{F,+}_k|d^{\dagger}_{k}d^{\dagger}_{-k}|\mu^{F,+}_k\rangle\nonumber\\
&+&|r^{-}_k|^{2}\langle \mu^{F,-}_k|d^{\dagger}_{k}d^{\dagger}_{-k}|\mu^{F,-}_k\rangle\Big],
\end{eqnarray}
where we replace the summation by integration. The ${\cal C}^{\infty}_{i,j}$ and ${\cal I}^{\infty}_{i,j}$ obtained so characterize a steady-state reduced density matrix $\rho_L$. The reduced density matrix is diagonalized in the orthogonal basis  $\{|\mu^{F,\pm}_{k}\rangle\}$ which can be described by a time-periodic generalized canonical ensemble~\cite{Apollaro2016,Pappalardi2017}. This state can be used to compute the Fisher information in Eq.~(\ref{eq:ssfisher}), as described in details in  Sec. B of the SM. 
%This $F^{ss}_{Q}$ in the steady-state using ${\cal C}^{\infty}_{i,j}$  and ${\cal I}^{\infty}_{i,j}$  can be obtained following  the method presented in the SM.

%%%%%%%%%%%%%%%%%%%%% FIG4 %%%%%%%%%%%%%%%%%%%%%%%%%%%%%%%%%%%%%%%%%%%%%%

\begin{figure}
\center
\includegraphics[height=0.22\textheight]{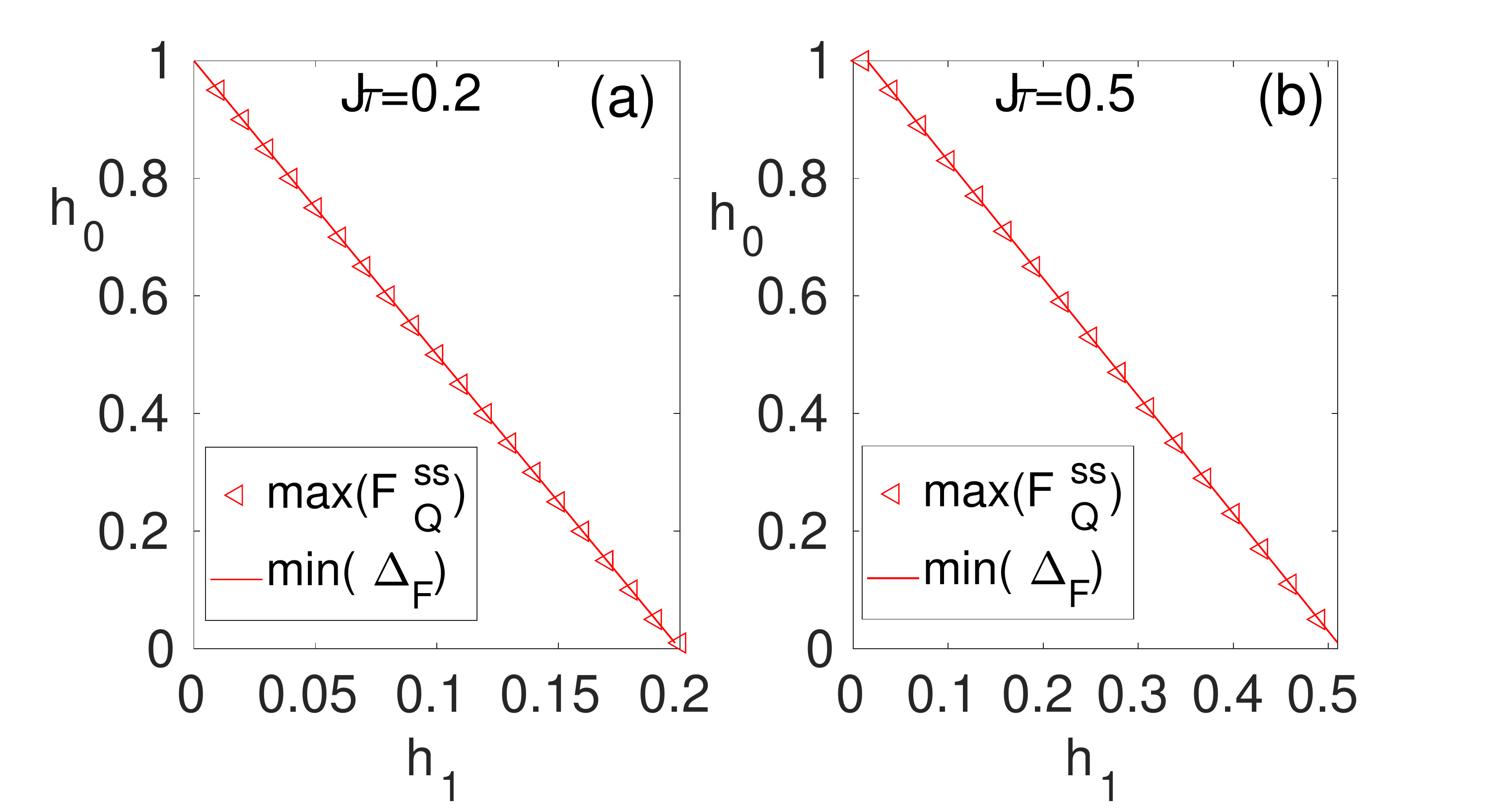}
\caption{The minimum of the Floquet gap $\Delta_F$ (red regular line) and the peak value of quantum Fisher information $F^{ss}_{Q}$ (red triangles) on the $h_0{-}h_1$ plane for: (a) $J\tau{=}0.2$; and (b) $J\tau{=}0.5$.  Here, the total system size is $N=2000$ and the block size is $L{=}4$.
}
\label{fig:fig4}
\end{figure}

One of the main advantages of our quantum-many body probes is the presence of another external parameter, namely the DC field $h_0$, which can be tuned to enhance the sensing precision.   
To see the effect of $h_0$ on the steady state QFI, in Figs.~\ref{fig:fig3}(a)-(d), we plot $F^{ss}_{Q}$ as a function of both $h_0$ and $h_1$ for different block sizes namely: (a) $L{=}1$; (b) $L{=}2$;  (c) $L{=}4$; and (d) $L{=}10$. As evident in the figures, by increasing the block size $L$, the $F^{ss}_{Q}$ increases considerably and peaks along a line in the plane of $h_0{-}h_1$.   It is shown in~\cite{Sen2016,Russomanno2016}, that the steady-state properties of periodically driven systems are closely linked to the spectrum of the Floquet Hamiltonian. Especially, it is shown that at the Floquet band crossing several peaks occur in the entanglement entropy.  To understand the origin of peaks in the $F^{ss}_{Q}$ in the present case, we fix $\tau$ and analyze the  Floquet gap $\Delta_{F}$ as a function of $h_1$ and $h_0$. The $\Delta_{F}$ is defined as 
\begin{eqnarray}
\Delta_{F} = \min_{k}(2|\vec{{\cal \mu}}^{F}_{k}|),  
\label{eq:flgap}
\end{eqnarray}
namely, the minimum gap between the two Floquet bands,   
$\mu^{F,+}=|\vec{{\cal \mu}}^{F}_{k}|$ and $\mu^{F,-}=-|\vec{{\cal \mu}}^{F}_{k}|$. The $|\vec{{\cal \mu}}^{F}_{k}|$ depends on $h_0,h_1$, and $\tau$. For a fixed $\tau,h_0$, and $h_1$, the minimum of $|\vec{{\cal \mu}}^{F}_{k}|$ occurs at $k=\pi$. Thus, $\vec{{\cal \mu}}^{i}_{k}$ becomes $\vec{{\cal \mu}}^{i}_{k=\pi} = (0, 0, h_0-J)$ which using Eq.~(\ref{eq:kfloquet}) gives $\cos(|\vec{{\cal \mu}}^F_{k=\pi}|\tau) = \cos((h_0-J)\tau+Jh_1)$.  For certain values of $h_0$ and $h_1$, it can be checked that $|\vec{{\cal \mu}}^{F}_{\pi}|=0$. Thus, for those value of $h_0$ and $h_1$, the Floquet band gap $\Delta_F=0$. By solving the former equation for $\vec{{\cal \mu}}^{F}_{k=\pi} =0$, we get \begin{eqnarray}
 Jh_1 = \tau|h_0-h_c|.
\label{eq:band_gap}
\end{eqnarray}
Interestingly, we find that for a fixed $\tau$ the peaks of $F^{ss}_{Q}$  occur along a straight line in the $h_0{-}h_1$ plane where the Floquet gap $\Delta_F$ vanishes.  
In Figs.~\ref{fig:fig4}(a)-(b) we plot the location of points in the $h_1{-}h_0$ plane where $\Delta_F$ is minimum and $F^{ss}_Q$ is maximum. The two lines perfectly collapse on each other showing that the vanishing Floquet gap corresponds to the maximum of the steady state QFI for various choices of $\tau$. This resembles the correspondence between the closing of the energy gap at the critical point and the maximization of the QFI in the ground state quantum sensing with global accessibility.

%%%%%%%%%%%%%%%%%%%%% FIG5 %%%%%%%%%%%%%%%%%%%%%%%%%%%%%%%%%%%%%%%%%%%%%%%%%%%%%%%%%%%%%%%%%%%%%%%%%
\begin{figure}
\centering
\includegraphics[height=0.3\textheight]{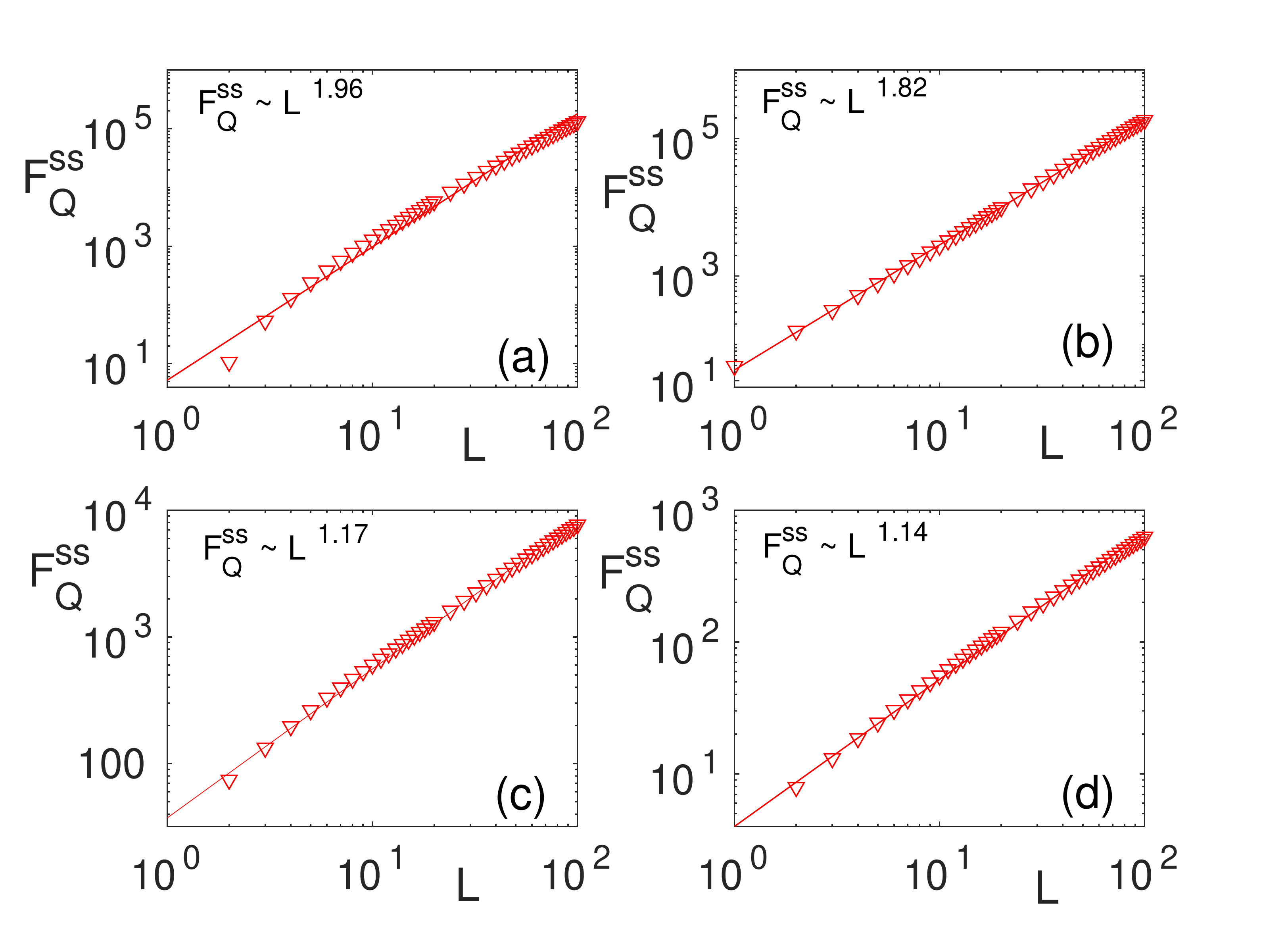}
%{Line_MAx_QFI_L_5_Zero_Floquet_gap}
\caption{Quantum Fisher information $F^{ss}_Q$ versus the block size $L$ in a system of length $N=10^4$ and time period of $J\tau=0.2$: (a) $(h_0/J,h_1/J) = (0.191,0.161)$;  (b) $(h_0/J,h_1/J) = (0.83,0.034)$; (c) $(h_0/J,h_1/J) = (0.161,0.191)$; and (d) $(h_0/J,h_1/J) = (0.6,0.2)$.  The plots in panels (a) and (b) belong to the vanishing Floquet gap line while the panels (c) and (d) are away from that. The triangles represent original numerical data while the solid red lines are the fitting curve using the least-square method.}
\label{fig:fig5}
\end{figure}

\section{ Steady state scaling of quantum Fisher information} 
One important features of quantum sensing in comparison with classical methods is resource efficiency. This is quantified through  scaling of the QFI with respect to the number of resources needed to perform the estimation.
In our setup, 
we have access to spins in a block of size $L$  which is explained by the density matrix $\rho_L$. Since all the measurements will be performed on this block, it is reasonable  to consider the number of spins $L$ as the resource for our quantum sensing protocol. To quantify the effectiveness of our steady state sensing protocol, one has to investigate the scaling of $F^{ss}_{Q}$ as  a function of resources $L$. Therefore, by fixing $h_0$ and $h_1$,  one can explore how $F^{ss}_{Q}$ (which is computed with respect to $h_1$) changes with increasing $L$. In particular, we fit the numerical data with the fitting function of the form $f(L){=}A L^{\eta}$ such that for every choice of pair $(h_0,h_1)$, one gets $F^{ss}_{Q}{\approx}f(L) $. In general, $A(h_0,h_1)$ and $\eta(h_0,h_1)$ are functions of $h_0$ and $h_1$.  The exponent $\eta{=}1$ corresponds to the classical standard limit and any $\eta{>}1$ shows quantum enhanced sensing, with $\eta{=}2$ being the Heisenberg limit.  In Fig.~\ref{fig:fig5}(a), we fix $(h_0/J,h_1/J) = (0.191,0.161)$ which corresponds to one point along the line with vanishing Floquet gap where the QFI is maximum. Surprisingly, by considering block sizes of  $L{=}1{-}100$, the steady state QFI shows  scaling with $F^{ss}_{Q}\sim L^{1.96}$, which is well beyond the standard limit. In  Fig.~\ref{fig:fig5}(b), we take  $(h_0,h_1){=}(0.83,0.034)$ as another point on the vanishing Floquet gap line where the fitting gives $F^{ss}_{Q} \sim L^{1.82}$, which again shows quantum enhanced sensing. For the sake of completeness, in  Figs.~\ref{fig:fig5}(c)-(d), we plot $F^{ss}_{Q}$ versus block size $L$ for the two representative pairs  of $(h_0,h_1){=} (0.161,0.191)$ and  $(h_0,h_1){=}(0.6,0.2)$ away from the vanishing Floquet gap line. Interestingly, for these choices, although $\eta$ still exceeds the standard limit, it is considerably smaller than the choices of the points on the vanishing Floquet gap line. These findings are the key results of this paper and are analogous to the enhanced sensitivity near the ground state critical point~\cite{invernizzi2008optimal,Zanardi2008}, where the energy gap of the system vanishes. As criticality is a resource for  ground state quantum sensing, the vanishing of the Floquet gap can also be considered a resource for steady state quantum metrology.

It is worth emphasizing that there is a fundamental difference between our protocol and the conventional criticality enhanced sensitivity in the ground state of many-body systems. In such scenarios, the Fisher information is computed for the whole system assuming global accessibility. In our case, while the whole system remains a pure state, the local subsystem becomes mixed due to entanglement with the rest of the system. Due to this mixedness, some information may get lost and sensing is more challenging. Nonetheless, our analysis shows that in integrable systems the local steady state still carries a wealth of information about the AC field allowing for sensitivity near the Heisenberg limit.  This is non-trivial as, for instance, in GHZ-based quantum sensing~\cite{Giovannetti2006,Cable2007} even losing one particle completely destroys the quantumness of the probe.    

We would like to mention that the scaling analysis carried out in this section is robust with the increase of the total system size $N$. We have considered $N$ in the range of $N=2000$ to $N=10000$ for which the value of the scaling exponent $\eta$ remains pretty much robust as shown in Figs.~\ref{fig:fig5}(a)-(d). Moreover, the scaling exponents have been extracted for $L=1{-}100$ as further increasing the block size $L$ hardly changes the fitting function and the exponent $\eta$.

%%%%%%%%%%%%%%%%%%%%% FIG6 %%%%%%%%%%%%%%%%%%%%%%%%%%%%%%%%%%%%%%%%%%%%%%%%%%%%%%%%%%%%%%%%%%%%%%%%%%
\begin{figure}
\center
	\includegraphics[height=0.22\textwidth]{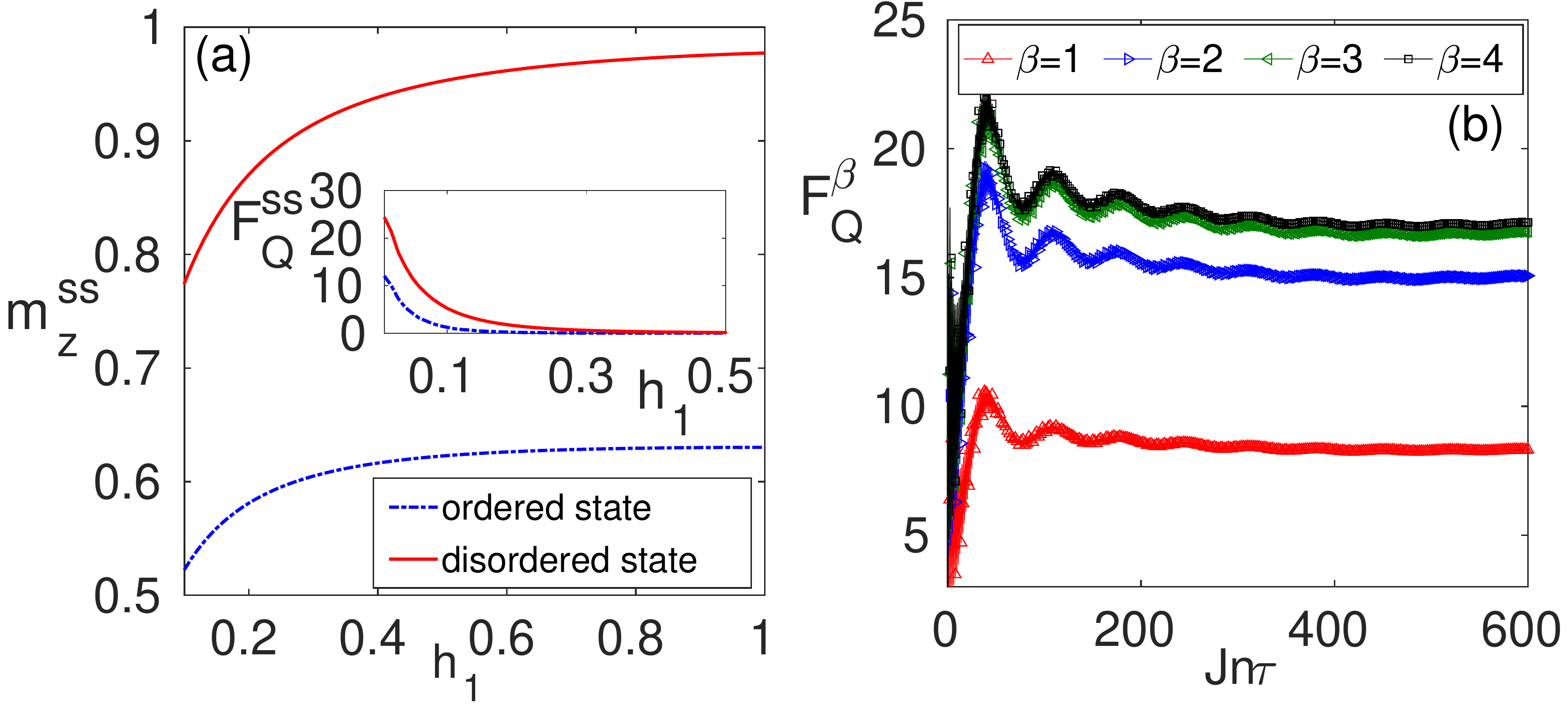}
	\caption{(a) The variation of the steady-state magnetization as a function of $h_1$ for disordered and ordered initial states, respectively. The corresponding steady-state QFI is shown in the inset. (b) The QFI  ($F^{\beta}_{Q}$) of a block of size $L=2$ as a function of time $t{=}n\tau$ for a thermal initial state at the finite temperature $T{=}\frac{1}{\kappa\beta}$. Here, $(h_0/J,h_1/J){=}(0.191,161)$, $J\tau{=}0.2$, and $N{=}2000$.}  
	\label{fig:fig6}
\end{figure}

%%%%%%%%%%%%%%%%%%%%% FIG7 %%%%%%%%%%%%%%%%%%%%%%%%%%%%%%%%%%%%%%%%%%%%%%%%%%%%%%%%%%%%%%%%%%%%%%%%%%
	
\begin{figure}
\centering
		\includegraphics[height=0.25\textheight]{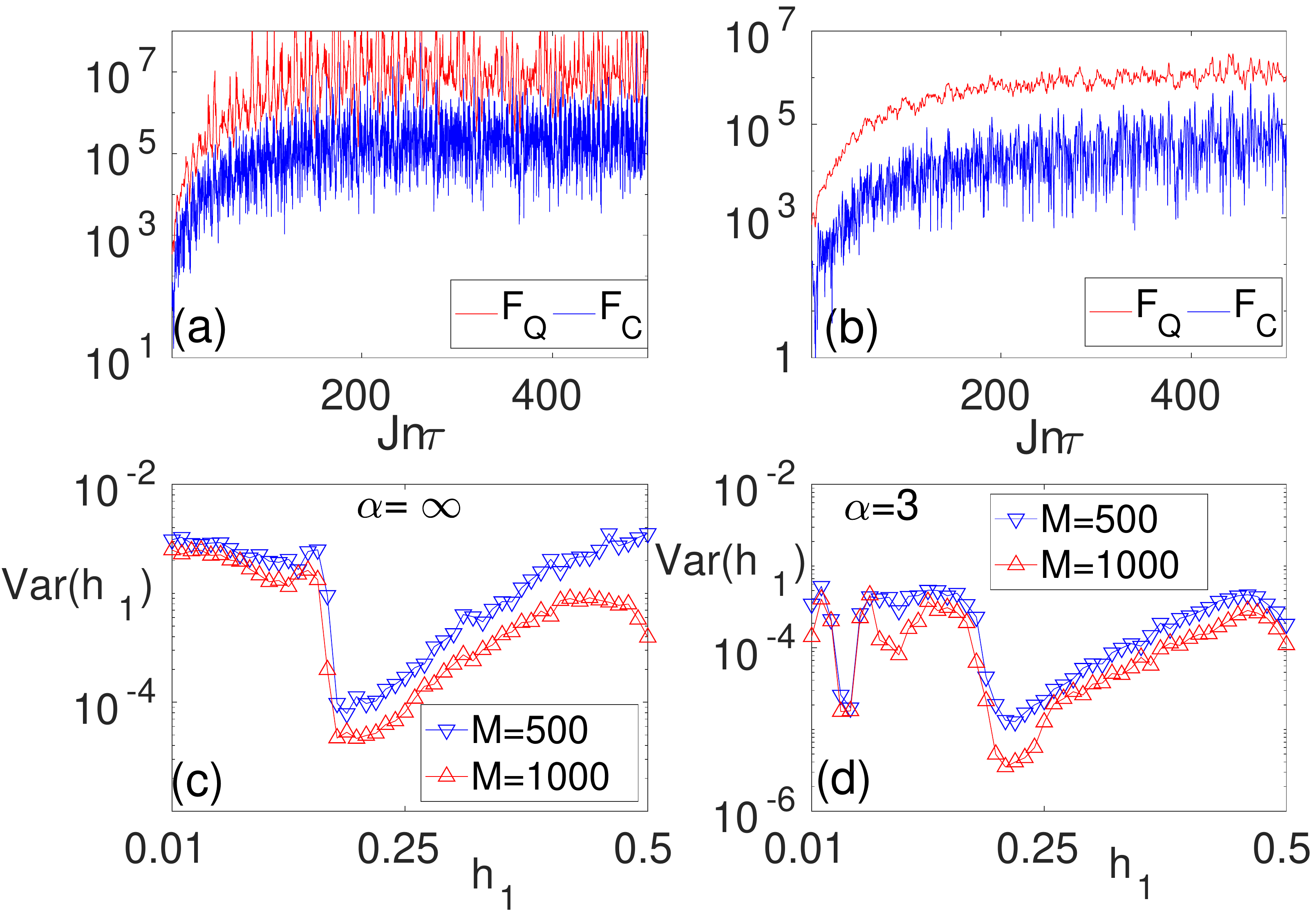}
		\caption{The evolution of $F_Q$ and $F_C$ as a function of time when $(h_0/J,h_1/J){=}(0.191,0.161)$ are tuned to be on the line of the vanishing Floquet gap for: (a) $\alpha \rightarrow \infty$; and (b) $\alpha=3$. The variance Var($h_1$) in the estimation of $h_1$ using Bayesian inference for two different numbers of repetitions $M$ for: (c) $\alpha \rightarrow \infty$; and (d) $\alpha=3$. The other parameters are $N{=}13$, $J\tau{=}0.2$, and $L=4$.} 
		\label{fig:fig7}	
\end{figure}

\section{Role of the initial state} 
In this section, we discuss  the role of the initial state for the estimation of $h_1$. For this, in Fig.~\ref{fig:fig6}(a), we plot the steady-state magnetization $m^{ss}_z$ as a function of $h_1$ for two different initial states, namely: (i) ordered state $|\Psi(0)\rangle{=}|\rightarrow\otimes\rightarrow \otimes\ldots\otimes\rightarrow\rangle$ (with $|\rightarrow\rangle=(|\uparrow\rangle+|\downarrow\rangle)/\sqrt{2}$); and (ii) disordered state $|\Psi(0)\rangle{=}|\uparrow\otimes\uparrow\otimes\ldots\otimes\uparrow\rangle$.  For both of these cases, the $m^{ss}_z$ starts from its initial value at $h_1{=}0$ and saturates for large $h_1$. The slope of $m^{ss}_z$ at any $h_1$ captures the degree of sensitivity for a small change on $h_1$, which in turn gives the information about $h_1$ that can be obtained from the measurement of $m^{ss}_z$.  In the inset of Fig.~\ref{fig:fig6}(a), we plot $F^{ss}_{Q}$ for the two different initial states. It is clear from the figure that the QFI takes larger values for the disordered initial state.

In order to consider a mixed initial state, we explore the performance of our AC field quantum sensing for a thermal initial state too. For this, the system is prepared initially in a thermal state  $\rho = e^{-\beta H_0}/{\cal Z}$, where ${\cal Z} = \mbox{Tr}( e^{-\beta H_0})$ is the normalization constant and $H_0$ is the time-independent Hamiltonian, namely the Hamiltonian in Eq.~(\ref{eq:model}) without the term $h(t)$. 
%\begin{eqnarray}
%H_0&=& - J \sum_{i=1}^{N}\hat{\sigma}^{x}_{i}\hat{\sigma}^{x}_{i+1} - h_0 \sum_{i}\hat{\sigma}^{z}_{i},
%\label{eq:model2}
%\end{eqnarray}
The subsequent dynamics can be obtained as $\rho(t){=}U(t)\rho U^{\dagger}(t)$, 
where $U(t)$ is a unitary operator given in Eq.~(\ref{eq:Evolution_U}).  We obtain a reduced density matrix between two-spins, i.e., $L=2$, and calculate the quantum Fisher information $F^{\beta}_{Q}$ as a function of time $t{=}n\tau$. In Fig.~\ref{fig:fig6}(b), we plot $F^{\beta}_{Q}$ as a function of time for different $\beta{=}1/\kappa T$, where $T$ is the temperature of the system and $\kappa$ is the Boltzmann constant. Here, we have taken the values of $h_0/J{=}0.191$ and $h_1/J{=}0.161$ which corresponds to the point where Floquet gap  $\Delta_F$ vanishes and $F^{ss}_{Q}$ shows a peak. From  Fig.~\ref{fig:fig6}(b), it can be seen that by increasing $\beta$ (decreasing temperature $T$), the $F^{\beta}_{Q}$ increases. Thus, we can infer that the uncertainty in the estimation of $h_1$ increases as the temperature increases. 
However, from Fig.~\ref{fig:fig6}(b) it is clear that $F^{\beta}_{Q}{\gg}1$, one can still get significant precision in the estimation of $h_1$ even at the finite temperature $\beta$.

\section{Realization on near-term quantum devices}
Near-term quantum devices are far from being perfect. They have several limitations in terms of the number of qubits, measurement types, and coherence time.  In addition, realizing a perfectly integrable system is challenging. We particularly, focus on ion trap systems in which the interaction between the qubits is described by the Hamiltonian~\cite{CMonroe_ion_trap, P_Zoller, R_Blatt,cold_ion_1,cold_ion_2}
\begin{eqnarray}
	H_{\alpha}(t)&=& -  \sum_{i,j}\frac{J}{r^{\alpha}_{ij}}\hat{\sigma}^{x}_{i}\hat{\sigma}^{x}_{j} - \sum_{i}(h_0 + h(t))\hat{\sigma}^{z}_{i},
	\label{eq:LR}
\end{eqnarray} 
where $\alpha$ determines the strength of interaction between sites $i$ and $j$ and can be tuned experimentally. The case of $\alpha=0$ describes a fully connected graph in which all qubits interact with each other equally. On the other hand, in the limit of $\alpha\to \infty$ one recovers the integrable Hamiltonian as in Eq.~(\ref{eq:model}). In general, for finite values of $\alpha$, the above Hamiltonian is non-integrable. However, as $\alpha$ increases the non-integrability becomes weaker such that for $\alpha>1$ system behaves more like the nearest neighbor Ising model. In typical ion trap experiments, $\alpha$ varies in the range $0.5\le \alpha\le 3$, the coupling strength $J$ is in the range $J\in[10^{2},10^{4}]$ Hz, and the coherence time $T_2 \ge 10^{-3}$ s \cite{ion_trap_1}. We consider a system of size $N=13$ with $\alpha=3$. As we will see, such small systems with $\alpha=3$, despite being non-integrable, still do not reach the infinite temperature thermal state for their subsystems. Therefore, one can still efficiently use them for steady state sensing within the coherence time of the system. 
	
Since the optimal measurement basis is complex and in general $h_1$ dependent, we suggest using the non-optimal but simple block magnetization measurement, described in the previous section. For such measurement, one can compute the classical Fisher information and compare it with the QFI.
In Figs.~\ref{fig:fig7}(a)-(b) we plot both the CFI and QFI as a function of time in a system of length $N=13$, $(h_0,h_1)=(0.191,0.161)$, and $L=4$ for: (a)  $\alpha \to \infty$; and (b) $\alpha=3$, respectively. 
Interestingly, despite being non-integrable, the system shows very large classical and quantum Fisher information. In addition, the system reaches its steady state around $nJ\tau=100$. For a typical exchange coupling of $J\sim 10$ KHz~\cite{ion_trap_2}, one needs a coherence time of $\sim 10$ ms. This is within the capability of current ion trap technologies which have achieved coherence time of $300$ ms (extendable to $2.1$ s with dynamical decoupling) \cite{long_coherence}.

Any quantum sensing protocol requires an estimation algorithm which uses the measured data for estimating the unknown parameter. Indeed, only by using an optimal estimation algorithm, together with optimal measurements, one can saturate the Cram\'er-Rao bound. Bayesian estimation is known to be the optimal estimator~\cite{Bayesian_1,Bayesian_2,Bayesian_3,Bayesian_4} for large data sets. In Sec. C of SM we presented optimal measurement basis for $L=1$ and $L=2$ for the Hamiltonian given in  Eq.~(\ref{eq:model}). The optimal measurement basis so obtained cannot be generalized to higher $L$  due to the complexity involved. Thus, we restore to a simple measurement which can be accessible in experiments. Consider block magnetization measurement which results in a data set of $M$ samples $\mathbf{d}=\{(O_k,n_k)\}$, in which any measurement outcome $O_k$ appears $n_k$ times (with $k=1,2,\cdots, L+1$) such that $\sum_k n_k=M$. The probability distribution of the unknown parameter $h_1$ is determined as 
\begin{equation}
	\mathbb{P}(h_1|\mathbf{d})=\frac{\mathbb{P}(\mathbf{d}|h_1)\mathbb{P}(h_1)}{\mathbb{P}(\mathbf{d})},
\end{equation}
where, $\mathbb{P}(h_1|\mathbf{d})$ is the posterior, $\mathbb{P}(\mathbf{d}|h_1)$ is the likelihood, $\mathbb{P}(h_1)$ is the prior probability distribution of $h_1$, and $\mathbb{P}(\mathbf{d})$ is the normalization factor to make the posterior a valid probability distribution. In the absence of prior information, one can consider $\mathbb{P}(h_1)$ to be a uniform distribution over the interval of interest. The likelihood can be computed as
\begin{equation}
\mathbb{P}(\mathbf{d}|h_1)=\binom{M}{n_1,n_2,\cdots,n_{L+1}} \prod_{k=1}^{L+1}(p_k)^{n_k},
\end{equation} 
where, $p_k$ is the probability of measuring outcome $O_k$. The estimated value $h_1^{est}$ is the point at which the posterior $\mathbb{P}(h_1|\mathbf{d})$ takes its maximum. By repeating the procedure one can estimate the variance $\text{Var}(h_1)$. Using block magnetization measurement, in Figs.~\ref{fig:fig7}(c)-(d), we plot the variance as a function of $h_1$ in a system of length $N=13$, block size $L=4$ for: (a) $\alpha \rightarrow \infty$; and (b)  $\alpha=3$.  The variance remains below $10^{-2}$ throughout the considered interval. As expected, by increasing the sample size $M$ the variance decreases.

\section{Effect of the total system size}
\label{sec:totalNeffect}
\begin{figure}
	\center
	\includegraphics[height=0.14\textheight]{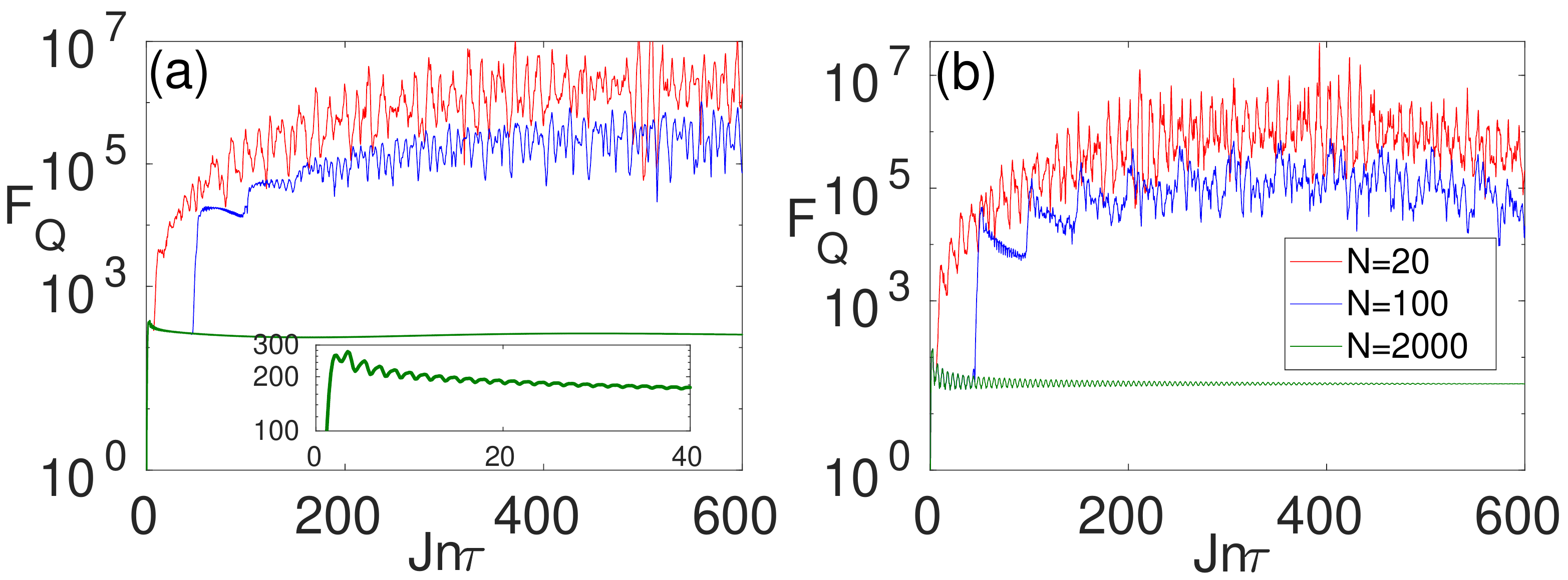}
	\caption{ Dynamics of $F_Q$ as a function of time $t{=}n\tau$ for various total system sizes and the time period $J\tau=0.2$: (a) $h_0/J{=}0.191,h_1/J{=}0.161$; and (b) $h_0/J{=}0.6,h_1/J{=}0.2$. Here $L=4$.The inset in (a) show the behavior of $F_Q$ in small time scale i.e., $t\approx 40/J$. }  
	\label{fig:fig8}
\end{figure}

So far, we have considered the situation in which the total system size is much larger than the subsystem of interest, namely $L\ll N$. This implies that the subsystem reaches its equilibrium and thus the reduced density matrix does not fluctuate in time which makes the sensing easier. However, current quantum devices are still very limited in terms of the number of qubits. Thus, it is important to see the performance of our protocol for fairly small total system sizes. In Figs.~\ref{fig:fig8}(a)-(b), we plot the $F_{Q}$ for a block of size $L=4$ and  different system sizes $N$ as a function of time $t=n\tau$ for: (a) $(h_0,h_1){=}(0.191,0.161)$; and (b) $(h_0,h_1){=}(0.6,0.2)$, respectively. The first choices of $h_0$ and $h_1$ are chosen along the peak of the $F^{ss}_{Q}$ whereas the second one away from the peak.  Interestingly, the QFI takes much larger values for the smaller system sizes which make sensing even more efficient. This is because in small systems the $L/N$ ratio is larger and there are fewer degrees of freedom over which the information is dispersed. As a result, the reduced density matrix $\rho_L$ contains more information about $h_1$ which reveals itself in larger values of the QFI. At the same time, since the total system size is smaller, the QFI shows more fluctuations for small systems which is a sign of a lack of full equilibration.  
	
%%%%%%%%%%%%%%%%%%%%% FIG8 %%%%%%%%%%%%%%%%%%%%%%%%%%%%%%%%%%%%%%%%%%%%%%

\section{Role of integrability}
The proposed protocol is very general and can be applied to any integrable time-independent Hamiltonian. In case of a non-integrable Hamiltonian, the periodic magnetic field leads to the heating phenomena \cite{DAlessio2014,Ishii2018}. Due to this heating effect, the long-time steady state is an infinite temperature state. Such an infinite temperature state no longer remains sensitive to the magnetic field $h_1$. Therefore, a non-integrable quantum sensor may not be useful for many-body steady state AC field quantum sensing.  
On the other hand, 
it is known that in integrable models under a periodic perturbation, the observable syncronize with the driving and do not heat up~\cite{Sen2016, Mishra2019}.

The dynamics of many-body system under periodic driving at the \textit{stroboscopic} time can be described by Floquet Hamiltonian $H^F$. For small $J\tau$, the Floquet Hamiltonian $H^F$ can be approximated by average Hamiltonian $H_{ave} =1/\tau \int_{0}^{\tau} H(t)dt$, i.e., $H^{F}\approx H_{ave}$.  For arbitrary $\tau$, the Floquet Hamiltonian is given by the Floquet-Magnus expansion. In Ref.~\cite{Ishii2018}, it is shown that  in generic integrable spin models the Floquet-Magnus expansion diverges around $J\tau \approx 1$, i.e. $H^F$ becomes infinite, which results in the sudden increase in the energy of the $H_{ave}$. This means that for $J\tau > 1$ even integrable systems can reach to infinite temperature state in their subsystems. For general Hamiltonians the sufficient condition for the convergence of Floquet-Magnus expansion is $J\tau \leq 1/||H(t)||$, where $||.||$ is the operator norm.  This means that in our model for $J\tau<1$, the long-time steady state is different from the infinite temperature state. It is this feature of integrability that is used in the present sensing protocol.

Note that the above argument does not necessarily mean that non-integrable systems cannot be used for steady state sensing at all. In fact, reaching the infinite temperature state requires large total system size $N$ and exponentially long time scales~\cite{Ishii2018,Rigol2019,NYYAO,VFine,Andreas2020}, in particular, if the non-integrability is weak. This very slow equilibration gives opportunity for quantum sensing before the system reaches the infinite temperature steady state. This is a crucial fact as, in practice, perfectly integrable systems might be difficult to realize. 
%Later in the paper, we will provide an example of such systems, based on long-range Ising model realizable in ion-traps, which is weakly non-integrable but still can be used efficiently  for steady state sensing.   	

 While, for the simplicity of the numerics, we mainly focus  on  the Dirac-delta AC field, the procedure is general and was used to infer the amplitude of a square AC field too (see the Sec. D of SM). We also discussed the main merit of our protocol over the existing one in Sec. E of SM.

%\section{Realization in near-term quantum simulators}

\section{Discussion and conclusion}
In this paper, we showed that the Ising model in a transverse field, as an integrable model, can be used for detecting the amplitude of an AC field. To enhance the precision of the estimation  a controllable DC transverse field is also applied.  By the combination of analytical and numerical simulation, based on Floquet formalism, we compute the quantum Fisher information of a block of spins when their reduced density matrix saturates to the steady state. We have four main results: (i)  in contrary to the conventional spin-echo and dynamical decoupling approaches, in which interaction between particles is not helpful, our approach harnesses such interactions for AC field sensing without demanding extra pulses; (ii) in clear distinction from the ground state critical sensing systems, our protocol only demands partial accessibility to the system; (iii) the steady state quantum Fisher information can reveal scaling beyond the standard limit, almost achieving the Heisenberg bound, with respect to the block size; and (iv) analytical analysis using the Floquet formalism, shows that this quantum enhanced scaling corresponds to the closing of the Floquet gap. Our results are general to all integrable systems in which Floquet heating does not occur. This means that the transverse Ising model can be used as a many-body sensor for all AC fields with $J\tau<1$.  However, we show that if the non-integrability is weak and the total system size is not very large, the non-integrable systems can still be used for efficient sensing too. Moreover, we have considered block magnetization as a simple, though sub-optimal, measurement basis that can be used in practice for efficient sensing. The resulting classical Fisher information is fairly close to the QFI, as the ultimate precision bound. Block magnetization measurement together with the Bayesian estimation algorithm have been used for quantum sensing to show the practicality of the protocol in near-term quantum simulators, such as the ion-traps.

\section*{ACKNOWLEDGMENTS}
A.B. acknowledges support from the National Key R \& D Program of China (Grant No.2018YFA0306703), the National Science Foundation of China (Grants No. 12050410253 and No. 92065115), and the Ministry of Science and Technology of China (Grant No. QNJ2021167001L). UM knowledges funding from the Chinese Postdoctoral Science Fund 2018M643437.

\section*{Author contributions statement}
A.B. conceived the preliminary idea. U.M. obtained the data and produced the figures. U.M. and A.B. equally contributed to the analysis and writing of the manuscript.

%\noindent LaTeX formats citations and references automatically using the bibliography records in your .bib file, which you can edit via the project menu. Use the cite command for an inline citation, e.g.  \cite{Hao:gidmaps:2014}.

%For data citations of datasets uploaded to e.g. \emph{figshare}, please use the \verb|howpublished| option in the bib entry to specify the platform and the link, as in the \verb|Hao:gidmaps:2014| example in the sample bibliography file.

\clearpage
\onecolumngrid
\begin{center}
{\large \bf Supplementary Material for  \protect \\ 
``Integrable Quantum Many-Body Sensors for AC Field Sensing'' }\\
\vspace*{0.3cm}
Utkarsh Mishra$^{1}$, and Abolfazl Bayat$^{1}$ \\
$^{1}${\small \em Institute of Fundamental and Frontier Sciences, University of Electronic Science and Technology of China, Chengdu 610051, China} \\
\end{center}

%%%%%%%%%% Prefix a "S" to all equations, figures, tables and reset the counter %%%%%%%%%%
\setcounter{equation}{0}
\setcounter{figure}{0}
\setcounter{table}{0}
\setcounter{page}{1}
\makeatletter
\renewcommand{\theequation}{S\arabic{equation}}
\renewcommand{\thefigure}{S\arabic{figure}}
\renewcommand{\bibnumfmt}[1]{[S#1]}
\renewcommand{\citenumfont}[1]{S#1}
%\newcommand{\section}[1]{\section}
%%%%%%%%%% Prefix a "S" to all equations, figures, tables and reset the counter %%%%%%%%%%

%\section{Appendix}
%%%%%%%%%%%%%%%%%%%%%%%%%%%%%%%%%%%%%%%%%%%%%%%%%%%%%%%%%%%%%%%%%%%%%%%%%%

\section{A. Floquet Hamiltonian}
\label{sec:method}
We present here the analytical description of the model
\begin{eqnarray}
H(t)=- J \sum_{i=1}^{N}\hat{\sigma}^{x}_{i}\hat{\sigma}^{x}_{i+1} - \sum_{i}(h_0 + h(t))\hat{\sigma}^{z}_{i},
\label{eq:smodel}
\end{eqnarray} 
where, $J>0$ is the nearest-neighbor spin-spin interaction, $h_0$ is a DC external magnetic field which is tunable,  $\hat{\sigma}_{i}^{x/y}$  are Pauli matrices at site $i$, and the periodic boundary conditions is assumed, i.e., $\hat{\sigma}_{N+1}^{x/y}=\hat{\sigma}_{1}^{x/y}$. The system represents interacting spin-1/2 particles with Ising type interaction on a $1D$ lattice of length $N$ in a transverse field to serve as a many-body probe for sensing a time-periodic magnetic field, $h(t)$. 
The Hamiltonian in Eq.~(\ref{eq:smodel}) can be solved analytically~\cite{Lieb1961,Barouch1971}. The first step is to map the spin operators, $\hat{\sigma}_{i}$, into fermionic operators, $\hat{c}^{\dagger}_{i}(\hat{c_{i}})$, defined via the following transformations:
\begin{eqnarray}
\hat{\sigma}^{-}_{j} &=& e^{i\pi\sum_{i=1}^{j-1}\hat{\sigma}^{+}_{i}\hat{\sigma}^{-}_{i}}\hat{c}_{j}\nonumber\\
\hat{\sigma}^{+}_{j} &=& \hat{c}^{\dagger}_{j}e^{-i\pi\sum_{i=1}^{j-1}\hat{\sigma}^{+}_{i}\hat{\sigma}^{-}_{i}},
\label{eq:JW1}
\end{eqnarray}
where, $\hat{\sigma}^{\pm}_{j} = (\hat{\sigma}^{x}_j\pm\hat{\sigma}^{y}_j)/2 $. The Hamiltonian as a result transformed into  a  quadratic-fermionic form
\begin{eqnarray}
H(t)&=& - J \sum_{i=1}^{N}(\hat{c}^{\dagger}_{j}\hat{c}^{\dagger}_{j+1} + \hat{c}_{j+1}\hat{c}_{j}+\hat{c}^{\dagger}_{j}\hat{c}_{j+1}+\hat{c}^{\dagger}_{j+1}\hat{c}_{j})\nonumber\\
&-& h(t) \sum_{i}(2\hat{c}^{\dagger}_{j}\hat{c}_{i}-1).
\label{eq:model1}
\end{eqnarray}
It is to be noted that the fermionic Hamiltonian in Eq.~(\ref{eq:model1}) is transnational invariant and therefore by applying a Fourier transformation  $\hat{c}_j = \frac{1}{\sqrt{N}}\sum_{k}e^{i k j} \hat{c}_k$, it can be written in  $k{-}$space. The full Hamiltonian can be decomposed into the sum of the Hamiltonian for each $k-$mode i.e.,  $H=\sum_{k}H_{k}$ with  $H_{k}$ being the Hamiltonian of the $k^{th}$ subspace given by $H_k = \sum_{k>0}(h(t)+J\cos(k))\Big(\hat{c}^{\dagger}_{k}\hat{c}_{k}-\hat{c}_{-k}\hat{c}^{\dagger}_{-k}\Big) + J\sin(k)(\hat{c}^{\dagger}\hat{c}^{\dagger}_{-k}-\hat{c}_{k}\hat{c}_{-k})$. It can be seen that each $H_k$ acts on a subspace spanned by basis $\{|0\rangle$, $c^{\dagger}_{k}c^{\dagger}_{-k}|0\rangle\}$, where $|0\rangle$ is the vacuum of the Jordan-Wigners fermions $\hat{c}_k$. By defining pseudo spin basis as $|\uparrow\rangle_k = |0\rangle$ and $|\downarrow\rangle_{k} = c^{\dagger}_{k}c^{\dagger}_{-k}|0\rangle$, we can write $H_k$ as
\begin{eqnarray}
H_{k} = \left(J \cos (k) + h(t)\right)\hat{\sigma}^{z}_p+J\sin (k) \hat{\sigma}^{y}_p,
\label{eq:kspace}
\end{eqnarray}
where $\hat{\sigma}^{y}_p$ and $\hat{\sigma}^{z}_p$ are pseudospin operators in the pseudospin basis and $k$ is termed as quasi-momentum which takes the values $k=\frac{\pi}{N},\frac{3\pi}{N}$,\ldots, $\frac{(N-1)\pi}{N}$ for even $N$. The Hamiltonian  in Eq.~(\ref{eq:model}) can be decomposed into the sum of even and odd parity-conserving Hamiltonians and  the ground state of the system belongs to the even parity subspace for every finite $N$ and it assumes BCS like form, given by~\cite{Damski2013} 
\begin{equation}
|\Psi(0)\rangle = \prod_{k>0}(u_{k}(0)+v_{k}(0)c^{\dagger}_{k}c^{\dagger}_{-k})|0\rangle,
\end{equation}
where $u_k = \sin(\theta/2)$, $v_k = \cos(\theta/2)$, and $\theta = \tan^{-1}\frac{J\sin(k)}{h_0 + J \cos(k)}$.  Thus, when $u_k = 1, v_k=0$ for all $k$, it corresponds to a state with all spins in the eigenbasis of $\hat{\sigma}^z$ with eigenvalue $+1$.  Under the dynamics given by the Hamiltonian in Eq.~(\ref{eq:model1}) the states in the two parity sectors  as well as those with different momentum evolves independently. Thus, for the unitary time dynamics, it is enough to consider states $\{|0\rangle$, $c^{\dagger}_{k}c^{\dagger}_{-k}|0\rangle\}$ which leads to time evolved state in the form~\cite{Apollaro2016}
\begin{equation}
|\Psi(t)\rangle = \prod_{k>0}(u_{k}(t)+v_{k}(t)c^{\dagger}_{k}c^{\dagger}_{-k})|0\rangle,
\end{equation} 
where $u_{k}(t)$ and $v_{k}(t)$ are the solutions of the Schr\"{o}dinger equation
\begin{eqnarray}
i\hbar \frac{d}{dt}(u_{k},v_{k})^{T} = H_{k}(t)(u_{k},v_{k})^{T}.
\label{eq:schrd}
\end{eqnarray}
The above analysis applies to any general time-dependent function $h(t)$. For stroboscopic dynamics, i.e., time-evolution of the system monitored in the steps $t = n\tau$, the state of the system at any time $t=n\tau$, can be obtained by repeated application of the unitary operator as
\begin{eqnarray}
|\Psi(t)\rangle = [U(\tau)]^{n}|\Psi(0)\rangle, 
\label{eq:state_evol}
\end{eqnarray}
where $U(\tau) = {\cal T}e^{-i \int_{0}^{\tau} H(t)dt}$,
is the the time-evolution operator for single time-period $\tau$ and ${\cal T}$ denotes time ordered product. 
For the Dirac-delta function, it is possible to find the effective Hamiltonian, known as Floquet Hamiltonian $H^F$,  which generates equivalent dynamics. Therefore, the equation for the ensuing dynamics, namely Eq.~(\ref{eq:state_evol}), can be simplified in term of the Floquet Hamiltonian as
\begin{eqnarray}
|\Psi(t=n\tau)\rangle = e^{-i n H^F \tau}|\Psi(0)\rangle
\nonumber\\
 =\prod_{k>0} e^{-inH^{F}_{k}\tau}|\psi_{k}(0)\rangle,
\end{eqnarray}
where $|\psi_{k}(0)\rangle = (u_{k}(0)+v_{k}(0)c^{\dagger}_{k}c^{\dagger}_{-k}))|0\rangle$.  Once the time-dependent state $|\Psi(t=n\tau)\rangle$ is known, the time-dependent magnetization at the stroboscopic time, $t=n\tau$, is expressed as $m_z(n\tau) = \langle\Psi(n\tau)|\frac{1}{N}\sum_{i=1}^{N}\hat{\sigma}^{z}_{i}|\Psi(n\tau)\rangle$. By employing the Jordan-Wigner transformation and expressing $\hat{\sigma}^z$ in terms of $\hat{\sigma}^{\pm}$, we have
\begin{eqnarray}
m_{z}(n\tau) = \frac{1}{N}\sum_{k>0}\Big(|u_{k}(n\tau)|^2-|v_{k}(n\tau|^2),
\end{eqnarray}
where, $u_{k}(n\tau)$ and $u_{k}(n\tau)$ are solutions of  the Schr\"{o}dinger equation given in Eq.~(\ref{eq:schrd}).

\section{B.  Quantum Fisher information of a block of size $L$}
\label{sec:QFIL}
%\noindent{\bf Quantum Fisher information of a block of size $L$:}
To calculate the quantum Fisher information of a subsystem of size $L$, we need to calculate the reduced density matrix of $L$ sites (we consider $L$ contiguous sites). The calculation of reduced density matrix requires partial tracing of complimentary degrees of freedoms.  For quadratic Hamiltonians of the form given in Eq.~(\ref{eq:model1}), this is accomplished by noting the relation  between the density matrix and the matrix of single-particle correlations of $L$ sites~\cite{Peschel_2009}. The form of the reduced density matrix which reproduces correct correlation matrix on $L$ sites is given by $\rho_L = e^{-\Omega}/Z$, where $\Omega$ is  quadratic in fermionic operators with energies $\epsilon_i$ i.e., $\Omega = \sum_{i=1}^{L}\epsilon_i \hat{c}^{\dagger}_{i}\hat{c}_{i}$ and $Z$ is the normalization constant. By breaking the complex fermions into Majorana basis defined as $\hat{c}_i = \frac{1}{2}(a_{2i-1}+ia_{2i})$ and $\hat{c}^{\dagger}_{i} = \frac{1}{2}(a_{2i-1}-ia_{2i})$, the density matrix of blocks of size $L$ can be represented as free fermionic  Gaussian state, given by
\begin{equation}
\rho_L = \frac{e^{-\frac{i}{4} \vec{a}^{T}\Omega \vec{a}}}{Z},
\label{eq:ferm_Gaussian}
\end{equation}
where, $Z = \mbox{Tr}[e^{-\frac{i}{4} \vec{a}^{T}\Omega \vec{a}}]$. Here $\Omega$ is a $2L\times 2L$ real antisymmetric matrix and $\vec{a}\equiv (a_1,\ldots,a_{2L})^T$ is a $2L$- dimensional array of Majorana fermions.  
The Gaussian state in Eq.~(\ref{eq:ferm_Gaussian}) is completely characterized by a two-point correlation matrix $\Gamma$ whose elements are given by
\begin{equation}
\Gamma_{ij} = \mbox{Tr}(\rho_L a_{i}a_j).
\end{equation}   
The $\Gamma$ matrix is an antisymmetric matrix. The elements of the $\Gamma$ matrix can be obtained in terms of ${\cal C}$ and ${\cal I}$ matrix where, $C_{i,j}=\mbox{Tr}[\rho_L \hat{c}^{\dagger}_i \hat{c}_j]$ and ${\cal I}_{i,j} = \mbox{Tr}[\rho_L \hat{c}^{\dagger}_i \hat{c}^{\dagger}_j]$. In terms of these matrices, we have~\cite{Yates2017}
\begin{eqnarray}
\Gamma_{2i-1,2j-1} &=& \delta_{i,j}+2i{\Im}[C_{i,j}+{\cal I}_{i,j}]\nonumber\\
\Gamma_{2i-1,2j} &=& i\delta_{i,j}-2i{\Re}[C_{i,j}-{\cal I}_{i,j}]\nonumber\\
\Gamma_{2i,2j-1} &=& -i\delta_{i,j}+2i{\Re}[C_{i,j}+{\cal I}_{i,j}]\nonumber\\
\Gamma_{2i,2j} &=& \delta_{i,j}+2i{\Im}[C_{i,j}-{\cal I}_{i,j}]\nonumber,
\end{eqnarray}
where, ${\Re}[\cdot]$ (${\Im}[\cdot]$) represents the  real (imaginary) part, $i,j=1,\ldots,L$, $\delta_{i,j}$ is discrete Kronecker delta function, and the elements of the time-dependent correlation matrix, $C(t)$, and anomalous correlation matrix, ${\cal F}(t)$, are given as
\begin{eqnarray}
C_{i,j} &=& \langle \Psi(t)|\hat{c}^{\dagger}_{i}\hat{c}_{j}|\Psi(t)\rangle \nonumber\\
        &=& \frac{2}{N}\sum_{k}|u_{k}(t)|^{2}\cos(k(j-i)),
\end{eqnarray}
and
\begin{eqnarray}
{\cal I}_{i,j} &=& \langle \Psi(t)|\hat{c}^{\dagger}_{i}\hat{c}^{\dagger}_{j}|\Psi(t)\rangle\nonumber\\
&=&   \frac{2i}{N}\sum_{k>0}u^{*}_{k}(t)v_{k}(t)\sin(k(j-i)).
\end{eqnarray}
Once the $\Gamma$ matrix is known, the quantities of interest can be expressed in terms of the $\Gamma$ matrix. For example, the symmetric logarithmic derivative has been obtained for the Gaussian states, Eq.~(\ref{eq:ferm_Gaussian}),  and in the Majorana basis it has the following form~\cite{Carollo2018}
\begin{eqnarray}
\hat{L}=\frac{1}{2}\vec{a}^T K \vec{a}+\vec{\zeta}^{T} \vec{a}
+\Lambda,
\end{eqnarray}
where $K$ is a Hermitian anti-symmetric matrix of dimension $2L\times2L$, $\zeta$ is a real vector and $\Lambda$ is a real number. The  matrix elements of the $K$ in the eigenbasis of  $\Gamma = \sum\limits_{r=1}^{2L}\gamma_{r}|\gamma_{r}\rangle\langle \gamma_{r}|$ are given by~\cite{Carollo2018}
\begin{eqnarray}
K_{rs}=\frac{(\partial_{h_1}\Gamma)_{rs}}{(\gamma_{r}\gamma_s-1)},
\end{eqnarray}
with $(\partial_{h_1}\Gamma)_{rs} = \langle r|\partial_{h_1} \Gamma|s\rangle$ and the partial derivative is taken with respect to the parameter to be estimated. By substituting $\hat{L}$ into $\mbox{Tr}[\rho_{L}\hat{L}^2]$, the QFI in the eigenbasis of $\Gamma$ is expressed as~\cite{Carollo2018,Carollo2019}

\begin{eqnarray}
F_Q = \sum\limits_{r,s=1}^{2L}\frac{\langle r|\partial_{h_1}\Gamma|s\rangle\langle s|\partial_{h_1}\Gamma|r\rangle}{(1-\gamma_r\gamma_s)}.
\label{eq:fihser_gamma}
\end{eqnarray}
We have used this expression to obtain the Fisher information of a block of size $L$ in our time-dependent model.  
The expression of QFI is valid for all parameters except when $\gamma_r = \gamma_s = \pm 1$.    At these values of $\gamma's$, the density matrix $\rho_L$ becomes singular and is not well defined.

\begin{figure}
\centering
	\includegraphics[height=0.3\textheight]{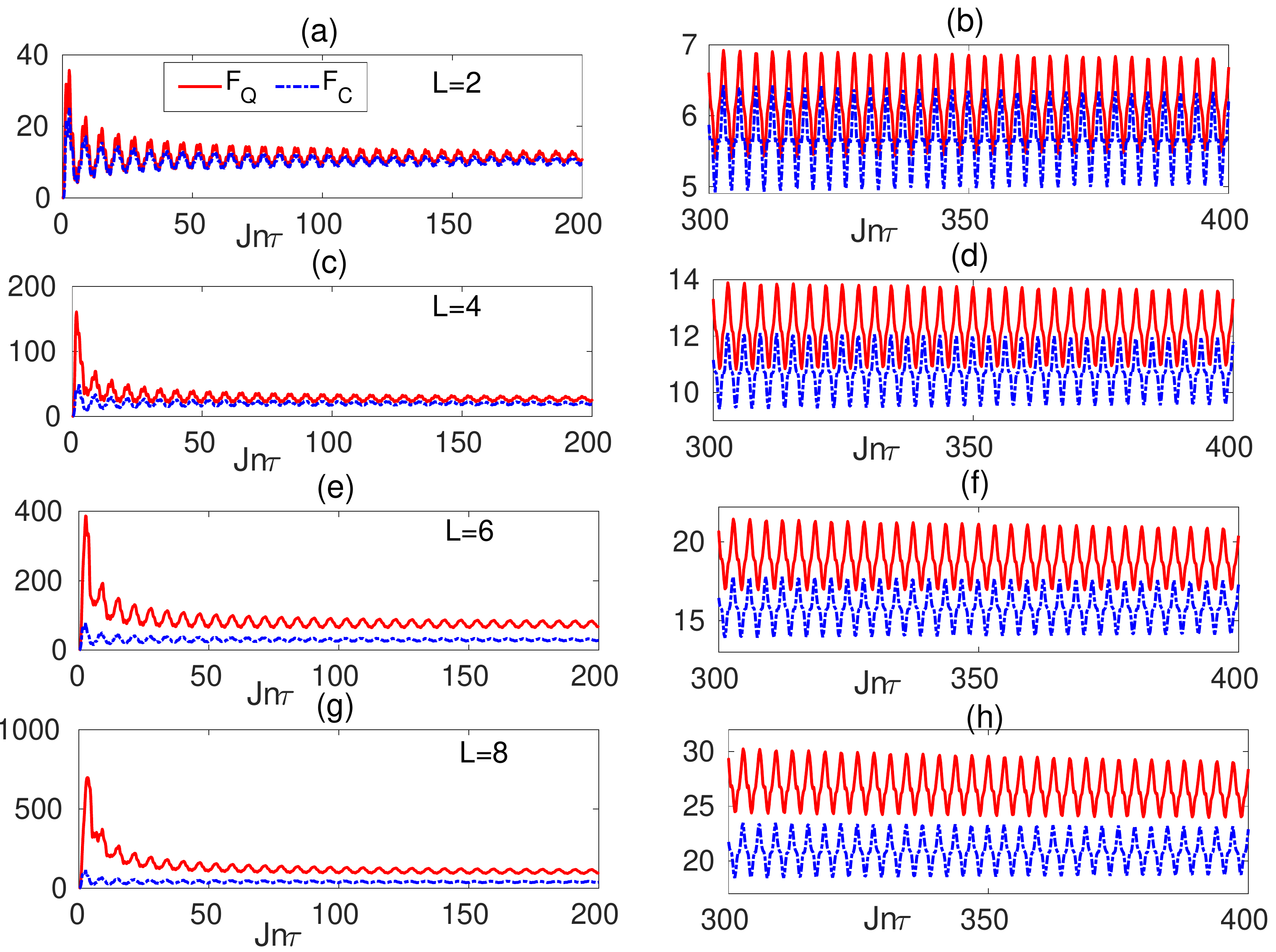}
	\caption{The QFI of a block of size $L{=}2,4,6,8$ is compared with the CFI resulted from the block magnetization measurement when the DC field is tuned at $h_0/J{=}1.0$. The plots are given for two different values of $h_1$, namely $h_1/J{=}0.1$ (the left panels) and $ h_1/J{=}0.2$ (the right panels). Here $J\tau=0.2$ and $N=2000$.
	}
	\label{fig:fig9}
\end{figure}
\section{C. Optimal versus sub-optimal measurements}
\label{sec:optimal}
In order to obtain the exact form of the optimal POVM, we first define the symmetric logarithmic derivative operator $\hat{L}$ to satisfy
\begin{eqnarray}
\partial_{h_1} \rho_L = \frac{\hat{L}\rho_L +\rho_L \hat{L}}{2}. 
\label{eq:SLD1}
\end{eqnarray}
By inserting the spectral decomposition form of $\rho_L$ in Eq.~(\ref{eq:SLD1}), one can easily obtain the following expression for $\hat{L}$  
\begin{eqnarray}
\hat{L} &=& \sum_{p}\frac{\partial_{h_1} \lambda_{p}}{\lambda_{p}} |\lambda_{p}\rangle \langle\lambda_{p}| \nonumber\\
&+& \sum_{p\neq q}\frac{\lambda_{p}-\lambda_{q}}{\lambda_{p}+\lambda_{q}} \langle \lambda_{p} |\partial_{h_1} \lambda_{q}\rangle|\lambda_{p}\rangle \langle\lambda_{q}|,\nonumber
\end{eqnarray}
where, $|\partial_{h_1} \lambda_{p}\rangle = \frac{\partial |\lambda_{p}\rangle}{\partial h_1}$.  The QFI then can be written as $F_Q(h_1) = \mbox{Tr}[\rho_L \hat{L}^{2}]$ and the eigenvectors of $\hat{L}$  provide the optimal POVM projectors $\{\Pi_{r} \}$~\cite{PARIS2009}. In general, the symmetric logarithmic derivative operator $\hat{L}$  and thus its eigenvectors, which are the optimal measurement basis,  depend on the unknown parameter $h_1$. Therefore, finding the optimal measurement basis is one of the big challenges in quantum estimation theory which often hinders saturating the  quantum Cram\'er-Rao bound. In fact, in most of the cases, the bound is only achievable  when sophisticated adaptive methods for updating the measurement basis are employed~\cite{bonato2016optimized,said2011nanoscale,higgins2007entanglement, berry2009perform,higgins2009demonstrating}.

Here we outline the calculation of symmetric logarithmic derivative for a single and two-qubit density matrix. A general single-qubit state of the system is written as $\rho_1 = \frac{1}{2}(\mathbb{I}+\vec{m}.\vec{\sigma})$, where $\vec{m} = \mbox{Tr}(\rho_1\vec{\sigma})$ and $\vec{\sigma} = (\hat{\sigma}^{x},\hat{\sigma}^{y},\hat{\sigma}^{z})$. It is to be noted that the Hamiltonian $H(t)$ is invariant under fermionic parity transformation, i.e., 
$H(t) = \Big(\prod\limits_{j=1}^{L}\hat{\sigma}^{z}\Big)H(t)\Big(\prod\limits_{j=1}^{L}\hat{\sigma}^{z}\Big)$. This implies, as shown in~\cite{Vidal2003,Russomanno2016}, that $m^{x},m^{y} =0$. Thus, we get a single-site density matrix which is diagonal in the eigenbasis of $\hat{\sigma}^{z}$ i.e., in $\{|\uparrow\rangle, |\downarrow\rangle\}$ basis. The symmetric logarithmic derivative in this basis is given by $\hat{L} = \frac{(\partial_{h_1} p)^2}{p}|\uparrow\rangle\langle \uparrow|+ \frac{(\partial_{h_1} (1-p))^2}{1-p}|\downarrow\rangle\langle \downarrow|$, where $p=(1+m_z)/2$. The  quantum Fisher information can be calculated using $F_{Q} = \mbox{Tr}[\rho_1\hat{L}^2]$ and it turns out as
\begin{eqnarray}
F_Q = \frac{(\partial_{h_1} m_z)^2}{(1+m_z)(1-m_z)}. 
\end{eqnarray}
On the other hand, the eigenvectors of $\hat{L}$ are  $\{|\uparrow\rangle, |\downarrow\rangle\}$.  If the set of POVM is constructed using the projections onto the eigenvectors of $\hat{L}$, then the classical Fisher information $F_C$ is given by 
\begin{eqnarray}
F_C & = & \frac{1}{p_{\uparrow}}(\langle \uparrow |\partial_{h_1} \rho_1|\uparrow\rangle )^2+ \frac{1}{p_{\downarrow}}(\langle \downarrow |\partial_{h_1} \rho_1|\downarrow\rangle)^2,
\end{eqnarray} 
where, $p_{\uparrow} = \langle \uparrow| \rho_1 |\uparrow\rangle$ and $p_{\downarrow} = \langle \downarrow| \rho_1 |\downarrow\rangle$.  A further simplification of $F_C$ gives $F_C = F_Q$. 

For the density matrix of two nearest-neighbor sites, one also needs to calculate the two-point correlators $\hat{\sigma}_{i}^{s}\otimes \hat{\sigma}_{i+1}^{s'}$($s,s' = x,y,z$). It can be seen that the correlators such as $\hat{\sigma}_{i}^{x}\otimes \hat{\sigma}_{i+1}^{z}$ and $\hat{\sigma}_{i}^{y}\otimes \hat{\sigma}_{i+1}^{z}$ vanishes due to  invariance of the Hamiltonian under parity transformation~\cite{Sen_De__2004,Mishra_2016}. %Moreover, in the long-time steady state, when the expectation value of the observables are time-independent, the correlations functions such as $\langle \hat{\sigma}_{i}^{x}\hat{\sigma}_{i+1}^{y}\rangle$ vanishingly small. 
Since periodic boundary conditions are assumed, the nearest-neighbor state is independent of which two neighboring sites are chosen for constructing the density matrix. The two-site density matrix of the system, therefore, is given by
\begin{equation}
\rho_2(t)=\begin{pmatrix} 
u_{11}& 0 &0 &u_{14} \\
0 & u_{22} &u_{23} &0 \\
0 & u_{32} & u_{33} &0 \\
u_{41}& 0 &0 &u_{44}
\end{pmatrix}
\end{equation}
where,
\begin{eqnarray}
u_{11} & = & \frac{1}{4} \left[ 1+2\langle \hat{\sigma}^{z}\rangle + \langle \hat{\sigma}_{i}^{z}\hat{\sigma}_{i+1}^{z}\rangle \right] \nonumber\\
u_{22} & = & u_{33} =\frac{1}{4}\left[ 1- \langle \hat{\sigma}_{i}^{z}\hat{\sigma}_{i+1}^{z}\rangle \right] \nonumber\\
u_{23} & = & \frac{1}{4} \left[ \langle \hat{\sigma}_{i}^{x}\hat{\sigma}_{i+1}^{x}\rangle+\langle \hat{\sigma}_{i}^{x}\hat{\sigma}_{i+1}^{y}\rangle+\langle \hat{\sigma}_{i}^{y}\hat{\sigma}_{i+1}^{x}\rangle 
+ \langle \hat{\sigma}_{i}^{y}\hat{\sigma}_{i+1}^{y}\rangle \right] \nonumber\\
u_{14} & = & \frac{1}{4} \left[ \langle \hat{\sigma}_{i}^{x}\hat{\sigma}_{i+1}^{x}\rangle -\langle \hat{\sigma}_{i}^{x}\hat{\sigma}_{i+1}^{y}\rangle - \langle \hat{\sigma}_{i}^{y}\hat{\sigma}_{i+1}^{x}\rangle 
- \langle \hat{\sigma}_{i}^{y}\hat{\sigma}_{i+1}^{y}\rangle \right] \nonumber\\
u_{44} & = & \frac{1}{4}\left[ 1 - 2\langle \hat{\sigma}^{z}\rangle + \langle \hat{\sigma}_{i}^{z}\hat{\sigma}_{i+1}^{z}\rangle \right].
\end{eqnarray} 
%\begin{eqnarray}
%u_{11} & = & \frac{1}{4}\left[1+2\langle \sigma^{z}\rangle(t) + \langle \hat{\sigma}_{i}^{z}\hat{\sigma}_{i+1}^{z}\rangle(t)\right]\nonumber\\
%u_{22} & = & \frac{1}{4}\left[1- \langle \hat{\sigma}_{i}^{z}\hat{\sigma}_{i+1}^{z}\rangle(t)\right]\nonumber\\
%u_{23} & = & \frac{1}{4} \left[ \langle \hat{\sigma}_{i}^{x}\hat{\sigma}_{i+1}^{x}\rangle(t)+\langle \hat{\sigma}_{i}^{x}\hat{\sigma}_{i+1}^{y}\rangle(t)+\langle \hat{\sigma}_{i}^{y}\hat{\sigma}_{i+1}^{x}\rangle(t) \nonumber\\
% &+& 
%\langle\hat{\sigma}_{i}^{y}\hat{\sigma}_{i+1}^{y}\rangle(t) \right] \nonumber\\
%u_{14} & = & \frac{1}{4}( \langle \hat{\sigma}_{i}^{x}\hat{\sigma}_{i+1}^{x}\rangle -\langle \hat{\sigma}_{i}^{x}\hat{\sigma}_{i+1}^{y}(t)\rangle - \langle \hat{\sigma}_{i}^{y}\hat{\sigma}_{i+1}^{x}(t)\rangle \nonumber\\&-& \langle \hat{\sigma}_{i}^{y}\hat{\sigma}_{i+1}^{y}(t)\rangle)\nonumber\\
%u_{44} & = & \frac{1}{4}(1 - 2\langle \hat{\sigma}^{z}(t)\rangle + \langle \hat{\sigma}_{i}^{z}\hat{\sigma}_{i+1}^{z}(t)\rangle)\nonumber\\
%\end{eqnarray}
The other non-zero elements are  given 
as $u_{32}=u^{*}_{23}$, and $u_{41}=u^{*}_{14}$. The non-zero correlators $\hat{\sigma}_{i}^{s}\otimes \hat{\sigma}_{i+1}^{s'}$ can be obtained using the formalism presented in~\cite{Mishra2019}. Once the two-site density matrix is obtained, the symmetric logarithmic derivative for the two-qubit state can be calculated. We  find that the symmetric logarithmic derivative with respect to the state $\rho_2(t)$ is given by

\begin{equation}
\hat{L}(t)=\begin{pmatrix} 
L_{11}& 0 &0 &L_{14} \\
0 & 1 &-1 &0 \\
0 & 1 & 1 &0 \\
L_{41}& 0 &0 &L_{44}
\end{pmatrix}.
\end{equation}
The eigenvectors of symmetric logarithmic derivative are given by 

\begin{eqnarray}
 |\ell_1\rangle &= & c_1(h_0,h_1)|\uparrow \uparrow\rangle +  c_2(h_0,h_1)|\downarrow \downarrow\rangle,
 \nonumber\\
  |\ell_2\rangle  & =& c_2(h_0,h_1)|\uparrow \uparrow\rangle +  c_3(h_0,h_1)| \downarrow  \downarrow \rangle, \nonumber\\
    |\ell_3\rangle &= & \left(|\uparrow  \downarrow\rangle - | \downarrow \uparrow\rangle\right)/\sqrt{2},
 \nonumber\\
  |\ell_4\rangle  & =& \left(|\uparrow  \downarrow\rangle + | \downarrow \uparrow\rangle\right)/\sqrt{2},
  \label{eq:SLD_2}
\end{eqnarray}
where $c_1 =  -\frac{-L_{11}+L_{44}+\sqrt{(L_{11}-L_{44})^{2}}+4L_{14}L_{41}}{2L_{14}L_{41}}$ and $c_2 = -\frac{-L_{11}+L_{44} - \sqrt{(L_{11}-L_{44})^{2}}+4L_{14}L_{41}}{2L_{14}L_{41}}$. As two of the eigenvectors namley $|\ell_1\rangle$ and $|\ell_2\rangle$ depends on the parameter to be estimated i.e., on $h_1$, the classical Fisher information obtained from the measurement of Eq.~(\ref{eq:SLD_2}) may not be equal to the quantum Fisher information~\cite{Liu_2019}.

The QFI is only a bound in the Cram\'er-Rao inequality and it is not guaranteed to be saturated unless an optimal measurement basis, as well as an optimal estimator, are chosen. As mentioned before, the measurement basis is determined by eigenvectors of the symmetric logarithmic derivative operator ${\hat L}$. In a general case, the optimal measurement basis depends on the sensing parameter $h_1$ which by definition is unknown making the optimal sensing impractical. Normally, to  overcome this problem, one has to update the measurement basis adaptively by extracting partial information about the unknown parameter using a sequence of non-optimal measurement basis~\cite{bonato2016optimized,said2011nanoscale,higgins2007entanglement, berry2009perform,higgins2009demonstrating}. Due to practical constraints, even if the optimal measurement basis is known (e.g. from an adaptive strategy)  its implementation may not be feasible. Therefore, one of the desired issues in quantum metrology problems is to find a suitable measurement basis that is close to the optimal one and is independent of $h_1$.

We consider a simple, though sub-optimal, measurement which is independent of $h_1$. The measurement is the block magnetization, which for a block of size $L$ takes $L+1$ outcomes from  $O_1=+L$ (when all the qubits are $|\uparrow\rangle$), $O_2=L-2$ (when except one qubit the rest are in the state $|\uparrow\rangle$) until $O_{L+1}=-L$ (when all the qubits are $|\downarrow\rangle$). 
Each of the outcomes $O_r$ appears with the probability $p_r$. Then one can  get the corresponding classical Fisher information $F_C$. Note that the block magnetization is easily measurable in ion traps~\cite{CMonroe_ion_trap, P_Zoller, R_Blatt} and superconducting quantum devices~\cite{Roushan_MBL,Guo_MBL,Ming_MBL}. 
In Figs.~\ref{fig:fig9}(a)-(h) we plot both the CFI, computed for the block magnetization, and the QFI as a function of time $t=n\tau$ for various block sizes.  In all these plots the control field is fixed to be $h_0/J=1$ while the left and the right panels represent the results for $h_1/J=0.1$ and  $h_1/J=0.2$, respectively. For the sake of clarity, the right panels are only shown for the later times. Interestingly, although the block magnetization is not the optimal measurement the resulted $F_C$ takes values greater than unity. This suggests that such a simple measurement can be used for efficient sensing.

%\section{Calculation of symmetric logarithmic derivative}
%\label{sec:SLD_A}

\section{D. Sensing Square pulses} 
\label{sec:square_pulse}
\begin{figure}
\centering
\includegraphics[height=0.3\textheight]{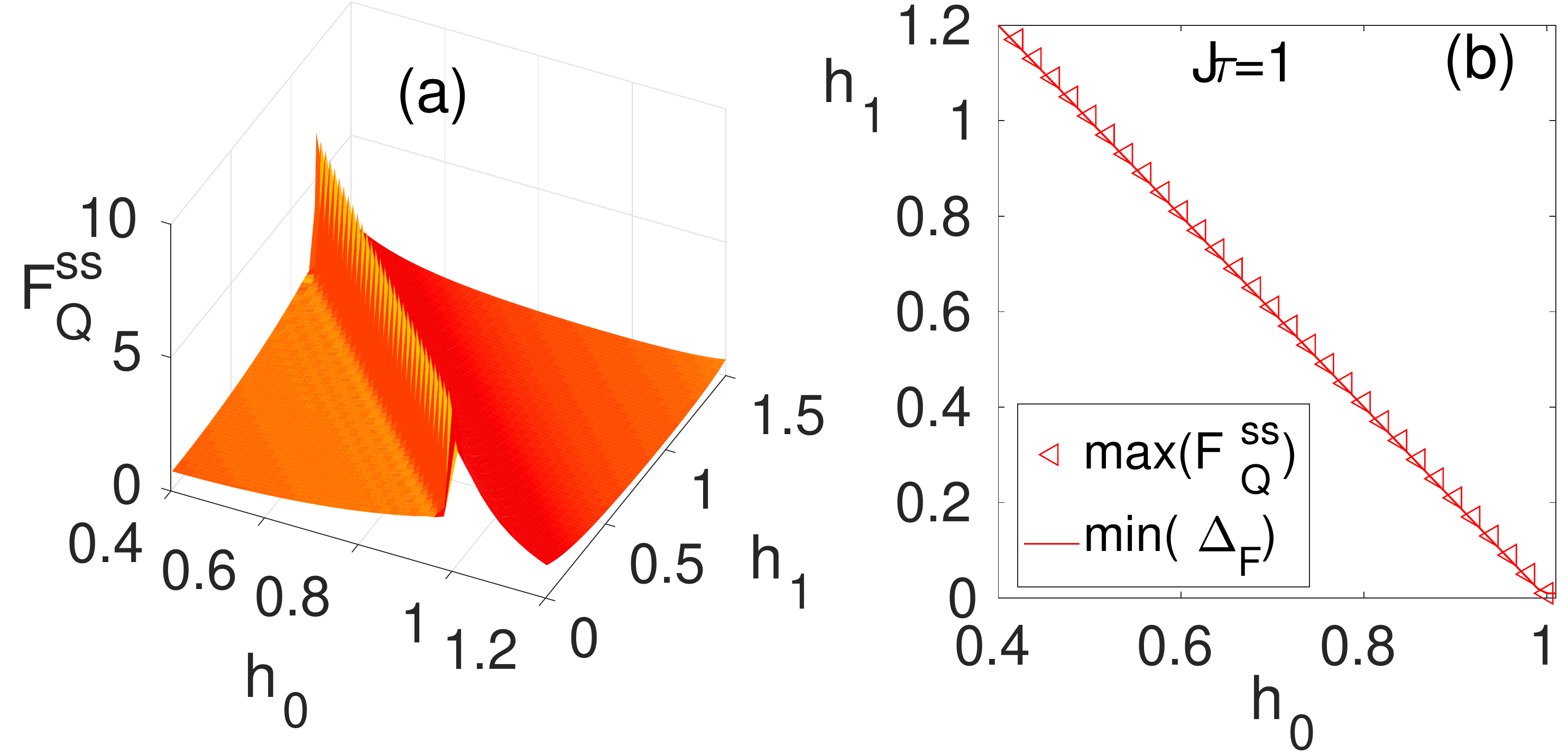}
\caption{(a) $F^{ss}_{Q}$ as a function of $h_0$ and $h_1$ for a square pulse magnetic field with $w=\tau/2$ for $L=4$. (b) For a block size of $L{=}4$, maximum of $F_Q^{ss}$ as a function of $h_0$ and $h_1$ is depicted by triangles and minimum of Floquet gap is shown with a regular line. Here, we fixed $J\tau{=}1$ and $N=2000$.
}
\label{fig:fig11}
\end{figure}

To see the generality of our approach, we also consider square pulsed form of the periodic field, given by the following equation  
\begin{equation}
 h(t) = \left\{ \begin{array}{ll}
         h_1 & \mbox{if $0 \leq t \leq w$};\\
        0 & \mbox{if $w \leq t\leq  \tau$},
        \end{array} \right. \
        \label{eq:square}
 \end{equation}
where $w$ characterizes the width of the pulse over an interval of $\tau$. 
The Floquet evolution operator over a time-period is given by 
\begin{eqnarray}
U(\tau) = e^{-iH_0 (\tau-w)}e^{-i (H_0+h_1\sum_{i=1}^{N}\sigma^{z}_i)w}. 
\end{eqnarray}
The local density matrix of the system reaches to a steady state under the AC field of the form given in Eq.~(\ref{eq:square}).
% For a fixed 
%$w=\tau/2$, 
%we plot the Floquet gap $\Delta_F$ for the square pulse in Fig.~(\ref{fig:fig8})(a). 
Similar to the case of Dirac-delta kick pulse, the Floquet gap takes its minimum along a straight line on the $h_0{-}h_1$ plane and $F^{ss}_{Q}$ exhibits peaks along the same line. In Fig.~\ref{fig:fig11}(a) we plot the steady state quantum Fisher information $F_Q^{ss}$ with respect to $h_0$ and $h_1$ for a block of size $L=4$.
The quantum Fisher information shows its peak exactly a straight line. In Fig.~\ref{fig:fig11}(b), we plot both max($F^{ss}_{Q}$) and min($\Delta_F$).  The lines  of max($F^{ss}_{Q}$) and  min($\Delta_F$) coincides. This shows the generality of the fact that the vanishing Floquet gap results in a higher sensitivity in steady state quantum metrology.

\section{E. Comparison with other protocols}
\label{sec:comparison}
Now in this section, we outline some of the key points about our protocol addressing its efficacy as compared to other existing protocols for AC-field sensing. We made this comparison with the two main existing schemes, namely spin echo and GHZ-based schemes. 

Spin echo utilizes a coherent superposition of spin states and a sequence of external pulses. The typical platform for realizing the spin echo mechanism for AC-field sensing is the nitrogen vacancy centers~\cite{Abraham1990,Mueller2014, Hansom2014}. However, the scheme is limited by the maximum time interval needed to accumulate phase and the quality of the coherent superposition of the spin states. To further enhance the precision, one can increase the number of spins. Although, it is crucial that the spins remain non-interacting as any interaction between them acts like decoherence and decreases the precision.  In fact, in the case of spin ensembles, the interaction is inevitable and one has to utilize a sophisticated pulse sequence and dynamical decoupling scheme~\cite{2019arXiv190710066Z}. Our proposal, however, takes a fundamentally different route as it exploits the interaction between the particles to drive the subsystems into a steady state. This naturally spares us from  any dynamical decoupling scheme. In addition, the enhancement in sensing precision near the vanishing Floquet gap is a resource that helps us to go beyond the standard limit. 

Another important feature of the proposed mechanism is its high precision performance despite demanding partial accessibility. Indeed, even with only $1{-}10$ percent accessibility one can still perform high precision sensing.  The importance of this becomes even more clear if one compares our protocol with the GHZ-based schemes~\cite{Giovannetti2006,Cable2007} in which even losing one particle completely loses its quantumness.

\end{document}